\documentclass[useAMS,usenatbib]{mn2e}
\usepackage{graphics,epsfig,psfig}
\usepackage[normalem]{ulem}
\usepackage{xcolor}
\usepackage{amsmath, amssymb}
\usepackage[]{inputenc,amssymb}

\def \be{\begin{equation}}
\def \ee{\end{equation}}
\def \bea{\begin{eqnarray}}
\def \eea{\end{eqnarray}}
\def \etal{{et al.}}

\definecolor{webgreen}{rgb}{0,.5,0}
\definecolor{webbrown}{rgb}{.6,0,0}
\usepackage[pdfpagelabels]{hyperref}
\hypersetup{%
   colorlinks=true,hyperfootnotes=false,%
   breaklinks=true,%
   plainpages=false, bookmarksnumbered, bookmarksopen=true,
   bookmarksopenlevel=1,%
   urlcolor=webbrown, linkcolor=webbrown, citecolor=webgreen,
   }

\title[Fermi Bubbles as Galactic wind]{Multi-wavelength features of Fermi Bubbles as signatures of a Galactic wind}

\author[Sarkar, Nath and Sharma]
{Kartick Chandra Sarkar$^{1,2}$ \thanks{kcsarkar@rri.res.in}, Biman B. Nath$^1$, Prateek Sharma$^2$\\
$^1$Raman Research Institute, Sadashiva Nagar, Bangalore 560080, India\\
$^2$Joint Astronomy Programme and Department of Physics, Indian Institute of Science, Bangalore 560012, India\\
}

\begin{document}
\maketitle

\label{firstpage}

\begin{abstract}
Using hydrodynamical simulations, we show for the first time  that an episode of star formation in the  centre of the Milky Way, with a star-formation-rate (SFR) $\sim 0.5$ M$_\odot$ yr$^{-1}$ for $\sim 30$ Myr, can produce bubbles that resemble the Fermi Bubbles (FBs), when viewed from the solar position. The morphology, extent and multi-wavelength observations of FBs, especially X-rays, constrain various physical parameters such as SFR, age, and the circumgalactic medium (CGM) density.
We show that the interaction of the CGM with the Galactic wind driven by star formation in the central region can explain the observed surface brightness and morphological features of X-rays associated with the Fermi Bubbles. Furthermore, assuming that cosmic ray electrons are accelerated {\it in situ} by shocks and/or turbulence, the brightness and morphology of gamma-ray emission and the microwave haze can be explained. The kinematics of the cold and warm clumps in our model also matches with recent observations of absorption lines through the bubbles.
\end{abstract}


\section{Introduction}

Galactic outflows are most easily seen in different wavebands in starburst galaxies such as M82 and NGC253 (e.g., \citealt{mccarthy1987}). However, the discovery of the gamma-ray emitting Fermi Bubbles (FBs) and the associated multi-wavelength features in the center of Milky Way (\citealt{su2010}) suggest that galactic outflows may be quite common, occurring in Milky-Way and lower mass galaxies frequently over their life-times. Supernova feedback is invoked to globally suppress star formation in halos less massive than $10^{12} M_\odot$ (\citealt{shankar06,sharma13}). However, it is well-known that isolated supernovae fizzle out in less than a Myr, requiring overlapping supernovae to 
form a superbubble (e.g., \citealt{sharma14}). Thus, galactic outflows driven by overlapping supernovae are likely to be ubiquitous.

In this paper we present a hydrodynamical model for explaining the multi-wavelength (gamma-rays, X-rays and radio) morphology and brightness of FBs. We assume a supernovae (SNe) driven Galactic wind as the driving mechanism for FBs.

Several observations point toward the existence of a gaseous outflow from the centre of Milky Way. An enhancement
in the diffuse soft X-ray emission in the longitude range $-20^\circ \le l \le 35^\circ$ with an emission scale height
(in the southern hemisphere) of $b \sim -17^\circ$ suggests a large-scale flow of gas out of
the disc \citep{snowden1995, everett2008}. This emission was modelled by \cite{snowden1995} with a mid plane 
gas density $n_e \sim 3.5 \times 10^{-3}$ cm$^{-3}$ and temperature $T\sim 4 \times 10^6$ K. Observations by
\cite{almy2000} proved that at least half of the central emission comes from more than 2 kpc from the Sun, and
most likely lies near the Galactic centre (see
also, \cite{park1997, yao2007}). \cite{almy2000} took into account other components (stellar, extragalactic),
and improved the model density and temperature to $n_e \sim 10^{-2}$ cm$^{-3}$ and $T \sim 8.2 \times 10^6$ K.
Interestingly, this emission was predicted from a model of cosmic ray driven Galactic outflow by \cite{breitschwerdt1994}. 
 In fact, using mid-infrared ($8.3 \mu$m) and ROSAT (1.5keV) observations, \cite{blandhawthorn2003} first showed the existence of a biconical Galactic outflow. They also speculated about the existence of projected x-ray bubbles on the both sides of the galactic plane extending up to $\sim 80^\circ$ in latitude.

The discovery of  $\gamma$-ray bubbles in the similar part of the sky, known as the Fermi Bubble, has given
spurt to exploring the high energetic implications of a Galactic outflow. These twin bubbles, extending up to $\sim 50^\circ$ ($\sim 8$ kpc in height) above and below the Galactic centre, are marked by $\gamma$-ray emission with a remarkably uniform surface brightness and a ($dN/dE \sim E^{-2}$) spectrum that is harder than the emission from the disc \citep{su2010}. 

The X-ray and $\gamma$-ray features also coincide with emission features in other wavelengths, such as the microwave haze found by WMAP and {\it Planck} \citep{finkbeiner2004, planck2013} and the polarized radio lobes seen at 2.3 GHz \citep{carretti2013}. Incidentally, \cite{lockman1984} had noted a HI hole in the inner Galaxy. These morphological similarities, to the extent of the edges of the features in different wavelength almost coinciding with each other, suggest a common physical origin.

Several models have been proposed to explain the FBs. As far as $\gamma$-ray emission mechanism is concerned, there remains an uncertainty whether the inverse Compton scattering of cosmic microwave background photons by  relativistic electrons is the source \citep{su2010} or the interactions of high energy protons with protons in the medium \citep{crocker2011}. The high energy electrons or protons can either be accelerated {\it in situ} by internal shocks and turbulence \citep{mertsch2011}, or advected from the disc. Outflows triggered by star formation in the Galactic centre (GC) region (\citealt{crocker2012,lacki14}) and by the black hole at the GC \citep{guo2012,yang12,mou14} have been proposed for the dynamical origin of the FBs.

The AGN-based models  (both jet and wind driven) generally consider a shorter age ($\lesssim$ few Myr) for the FBs because the inverse-Compton cooling time (due to up-scattering of starlight) for $\sim 100$ GeV electrons responsible for gamma ray emission is a few Myr (\citealt{guo2012,yang12}). The speed required to reach $\sim 10$ kpc in 1 Myr is $\sim 10^4$ km s$^{-1}$, achievable by relativistic jets slowed down by the hot circumgalactic medium (CGM). The power required for inflating young FBs is much higher and the outer shock is much stronger (\citealt{guo2012,zubovas12}), and X-ray emissivity and temperature much higher than what is observed (\citealt{kataoka2013}).  The SNe-driven models of FBs consider them to be long lived ($\gtrsim 10$ Myr), and thus the injected power is smaller. In fact, \citet{crocker2014b} suggest a hadronic origin for gamma ray emission and consider the FBs to be steady features older than few 100 Myr. The outer shock is weaker in the SNe-driven models and the temperature and emissivity jumps are modest, consistent with X-ray observations.

While the AGN jet and wind models have been explored numerically, simulations of a SNe driven model for FBs have not yet been carried out. The dynamical modelling is limited to simple arguments invoking a steady wind, termination shock, thermal instabilities, etc. (\citealt{crocker2014b,lacki14}). A realistic SNe-driven wind is expected to be affected by disc stratification and the presence of a CGM. Moreover, thermal and Rayleigh-Taylor instabilities are expected to mix the hot bubble gas with the halo gas. We capture these complex, time-dependent, multidimensional effects in the hydrodynamic numerical simulations presented in this paper.

 In our model, the FB is a time dependent phenomenon and is currently expanding. 
Our goal is to study the time dependent signatures of a star formation triggered Galactic wind, and to identify various features observed in different wavebands ($\gamma$-rays, X-rays, microwave  and radio) with various structural features of a Galactic wind. In doing so, we pay particular attention to  projection effects from the vantage point of the solar system in the Galactic disc. 

The paper is organised as follows. In section \ref{sec:sim}, we discuss the initial set up, the simulation settings and the parameters. The morphology and the importance of projection effects in the observations of the FBs  are described in section \ref{sec:res-morph}. In section \ref{sec:res-emission}, we describe the  X-ray, microwave and $\gamma$-ray emission from the FB. Section \ref{sec:kinematics} points out some kinematics aspects of the cold/warm clumps. We discuss the implications and improvements of our work in section \ref{sec:discussion}. Finally, in section \ref{summary}, we summarise the main conclusions of this paper.

\section{Simulation}
\label{sec:sim}
\subsection{Initial set up}
\label{subsec:setup}
The details of the initial set up for milky way (MW) type galaxies are given in a previous paper \cite{sarkar15}. However, for the sake of completeness, we briefly discuss the set up below. 

In our set up, we consider two gas components, a warm component ($T = 4\times 10^4$ K, including the contribution from non-thermal pressure) representing the disc gas, and a hot component ($T = 2.5 \times 10^6$ K) representing the extended circum-galactic medium (CGM). Since the warm gas represents the disc, we also consider azimuthal rotation for this component. The hot CGM, however, is considered to be non-rotating.

These two gas components are considered to be in steady state equilibrium with background gravitational potential of the stellar disc and dark matter (DM). For the disc, we use the Miyamoto \& Nagai potential \citep{miyamoto75}
\begin{equation}
 \Phi_{\rm disc }(R,z) = - \frac{GM_{\rm disc }}{\sqrt{R^2+(a+\sqrt{z^2+b^2}\,)^2}}, \:\: (\, a,b \geq 0 \,)  \,
\end{equation}
where $a$ and $b$ are the model parameters representing the scale length and the scale height of a disc of mass $M_{\rm disc }$ respectively, and, $R$ and $z$ are the cylindrical coordinates. For the dark matter, we use a modified form of NFW profile \citep{nfw96} introducing a core at the center. The modified form of the potential is given as
\begin{equation}
 \Phi_{\rm DM}(R,z) = - \left( \frac{GM_{\rm vir}}{f(c)\,r_s} \right) \frac{\log(1+\sqrt{R^2+z^2+d^2}/r_s)}{\sqrt{R^2+z^2+d^2}/r_s} \:\: (\, d \geq 0 ),
\end{equation}
where $ f(c) = \log(1+c)-c/(1+c) $ with $ c = r_{\rm vir}/r_s $ as the concentration parameter and $d$ is the radius of the core which gives a finite DM density at the centre.  $r_{\rm vir}$ and $r_s$ are, respectively, the virial radius and scale radius for a DM halo of mass $M_{\rm vir}$.

The steady state density distribution in a combined potential $\Phi (R,z) = \Phi_{\rm disc}(R,z) + \Phi_{\rm DM} (R,z)$ for the warm gas can be written as
\begin{eqnarray}
 \rho_{d}(R,z) &= & \rho_{d}(0,0)\,\exp \Bigl(-\frac{1}{c^2_{\rm sd}} \big[ \Phi(R,z)-\Phi(0,0) \nonumber\\
 && - f^2(\Phi(R,0)-\Phi(0,0))  \big] \Bigr) \,\,, 
\end{eqnarray}
and for the hot CGM,
\begin{equation}
 \rho_{h}(R,z) = \rho_{h}(0,0)\, \exp\left(-\frac{1}{c^2_{\rm sh}}\big[ \Phi(R,z)-\Phi(0,0)\big] \right)\,,
\end{equation}
where, $\rho_{d}(0,0)$ and $\rho_{h}(0,0)$ are the warm and hot gas central densities and 
$c_{\rm sd}$ and $c_{\rm sh}$ are the isothermal sound speeds of the warm disc and the hot CGM, respectively. Here, $f$ is the ratio of the disc gas rotation velocity and the stellar rotation velocity at any $R$ and taken to be a constant ($= 0.95$).  The density of a given location is, therefore, $\rho_d + \rho_h$. A full list of model parameters is given in Table \ref{table:model_para}.
 \begin{table}
  \centering
  \begin{tabular}{ l c l} 
   \hline\hline 
  parameters & values\\[0.5ex]
   \hline
   $M_{\rm vir} ({\rm M}_{\odot}) $      & $10^{12}$\\ 
   $M_{\rm disc } ({\rm M}_{\odot})$   & $5 \times 10^{10}$ \\
   $T_{\rm halo}$ (K) 			& $2.5 \times 10^6$ \\
   $r_{\rm vir}$ (kpc)                 & $258$ \\
   $c$                                 & $12$  \\ 
   $a$ (kpc)                            & $4.0$ \\
   $b$ (kpc)                            & $0.4$ \\ 
   $d$ (kpc)                             & $6.0$ \\
   $\mathcal{Z}_{\rm disc }$ ($Z_{\odot}$)   & $1.0$ \\
   $\mathcal{Z}_{\rm halo}$ ($Z_{\odot}$)   & $0.1$ \\
   $\rho_{d}(0,0)$ (m$_p$cm$^{-3}$)     & $3.0$\\
  \hline
   $\rho_{h}(0,0)$ (m$_p$cm$^{-3}$)   & $2.2 \times 10^{-3}$ \\
    \hline
  \end{tabular}
     \caption{Parameters used in our simulations. Hot gas central density $\rho_{h}(0,0)$ is obtained after normalising the total baryonic content (stellar plus gaseous) to $0.16$ of $M_{\rm vir}$, consistent with the cosmic baryonic fraction. While exploring the parameter space, we make this assumption flexible.}
 \label{table:model_para}
 \end{table}
\subsection{Code settings}
We use the publicly available code PLUTO \citep{mignone07} for our hydrodynamic simulations. We perform the simulations in 2D spherical coordinates assuming axi-symmetry around $\theta = 0$. The simulation box extends from $r_{\rm min} = 20$ pc to $r_{\rm max} = 15$ kpc in radial direction using logarithmic grids and from $\theta = 0$ to $\theta = {\rm \pi}/2$ in theta direction using uniform grids. This implies  that the disc lies on the $\theta = {\rm \pi}/2$ plane and our simulation box includes the first quadrant of the 2D slice taken along the $\theta$ plane of our Galaxy.

In our simulation, we express the temperature as $T \sim p/\rho$ which includes the hot gas pressure in addition to the $4\times 10^4$ K gas pressure inside the disc. The effective temperature of the disc is large enough to induce strong cooling unlike the warm gas at $T = 10^4$ K gas. In reality the disc gas is always being heated by the supernovae and other processes.  Since we are interested in the Galactic wind and not the disc ISM, we constrain the cooling of the disc material (but not the injected material) to be zero for a height less than $1.2$ kpc above the disc plane. A more detailed description about the code implementation can be found in \cite{sarkar15}.

\subsection{Injection parameters}
The mechanical luminosity of a starburst activity can be written as 
\begin{equation}
\mathcal{L} \approx 10^{40} \, {\rm erg\,s}^{-1} \, \epsilon_{0.3} \Bigl ( {{\rm SFR} \over 0.1 \, {\rm M}_\odot \, {\rm yr}^{-1}} \Bigr ) \,,
\label{eq:mech_en}
\end{equation}
where, $\epsilon_{0.3}$ is the thermalisation efficiency in units of $0.3$ and SFR is the star formation rate. Here we have considered 
Kroupa/Chabrier mass function, for which there is $\sim 1$ SN for every $100$ M$_\odot$ of stars formed. 

As we show later, the morphology and X-ray emission properties of FBs depend mostly on the combination of $\mathcal{L}$ and the CGM gas density. After scanning through various combinations of these two parameters, we show later (in \S 4.1, Figure 4) that a fiducial combination of $\mathcal{L}=5 \times 10^{40}$ erg s$^{-1}$ and $\rho_{h0} = 2.2 \times 10^{-3}$ cm$^{-3}$ best matches the observations. The implied star formation rate, according to eqn \ref{eq:mech_en}, is $\sim 0.5$ M$_\odot$ yr$^{-1}$ (considering $\epsilon_{0.3} = 1$). The current rate of star formation in the central molecular zone of Milky Way is of order $0.1$ M$_\odot$ yr$^{-1}$. Mid-infrared observations by \cite{yusuf-zadeh2009} have led to an estimate of SFR ranging between $0.007\hbox{--}0.14$ M$_{\odot}$ yr$^{-1}$, over the last 10 Gyr. Observations of young stellar objects in the central molecular zone (CMZ) in the 5-38 $\mu$m band with {\it Spitzer} allowed \cite{immer2011} to estimate a SFR of $\sim 0.08$ M$_{\odot}$ yr$^{-1}$. The diffuse hard X-ray emission in the Galactic centre region was used by \cite{muno2004} to estimate an energy input of $\sim 10^{40}$ erg s$^{-1}$. However, the star formation activity in the central region of the Galaxy is likely to be episodic.  Our fiducial SFR, averaged over the last several tens of Myr,  is therefore not unreasonable although it is a few times larger than the current SFR.

The mass injection rate has been taken as \citep{leitherer99}
\begin{equation}
\dot{M}_{\rm inj} = 0.3\, {\rm SFR}\,.
\end{equation}
In our fiducial simulation, the considered mechanical luminosity, $\mathcal{L} = 5 \times 10^{40}$ erg s$^{-1}$, corresponds to SFR $=0.5$ M$_{\odot}$ yr$^{-1}$ and therefore $\dot{M}_{\rm inj}=0.15$ M$_{\odot}$ yr$^{-1}$.

We inject this mass and energy in density and energy equations inside a region of $r \le r_{\rm inj}$ (60 pc). The injection rates can then be written as
\begin{equation}
\dot{p} = \frac{2}{3}\, \frac{\mathcal{L}}{(4\pi/3)\, r_{\rm inj}^3} 
\end{equation}
and
\begin{equation}
\dot{\rho} = \frac{\dot{M}_{\rm inj}}{(4\pi/3)\, r_{\rm inj}^3} \,\,,
\end{equation}
where, $p$ is the pressure. A full list of all the runs is given in Table \ref{table:runs}.

 \begin{table}
  \centering
  \begin{tabular}{ l c l} 
   \hline\hline 
  Name & $\mathcal{L}$ (erg s$^{-1}$) & $\rho_{h0}$ (m$_p$ cm$^{-3}$) \\[0.5ex]
   \hline
 S1	& $1.0 \times 10^{40}$ 	& $0.5 \times 10^{-3}$ \\
 S2	& $1.0 \times 10^{40}$	& $1.0 \times 10^{-3}$		\\
 S3	& $1.0 \times 10^{40}$	& $3.0\times 10^{-3}$ \\
 S4	& $2.0 \times 10^{40}$	& $0.5 \times 10^{-3}$ \\
 S5	& $2.0 \times 10^{40}$	& $1.0 \times 10^{-3}$ \\
 S6	& $2.0 \times 10^{40}$	& $2.0 \times 10^{-3}$ \\
 S7	& $2.0 \times 10^{40}$	& $3.0 \times 10^{-3}$ \\
 S8	& $4.0 \times 10^{40}$	& $0.7 \times 10^{-3}$ \\
 S9	& $5.0 \times 10^{40}$	& $1.1 \times 10^{-3}$ \\
 S10$^{\ast}$	& $5.0 \times 10^{40}$	& $2.2 \times 10^{-3}$ \\
 S11	& $6.0 \times 10^{40}$	& $3.0 \times 10^{-3}$ \\
 S12	& $1.0 \times 10^{41}$	& $1.1 \times 10^{-3}$ \\
 S13	& $1.0 \times 10^{41}$	& $2.2 \times 10^{-3}$ \\
 \hline
  \end{tabular}
     \caption{The list of runs  showing the injected mechanical luminosity and the central density in column 2 and 3, respectively. The fiducial run (S10) has been pointed out by an asterisk in the list. }
 \label{table:runs}
 \end{table}

\begin{figure}
 \includegraphics[trim=5.0cm 2.0cm 1.0cm 0.0cm, clip=true, width=6cm, height=10cm, angle=-90]{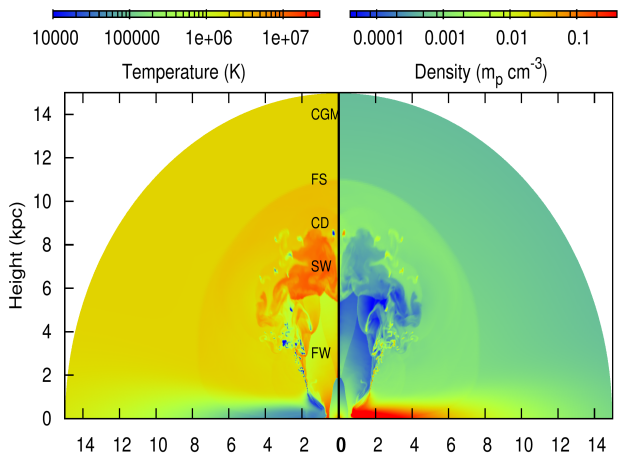}
 \caption {Snapshot of density (right panel) and temperature (left panel) contours at 27 Myr for our fiducial run (S10). The wind structure has been pointed out by different labels, from outside to inside as, CGM: circumgalactic medium, FS: forward shock, CD: contact discontinuity, SW: shocked wind and FW: free wind.}
 \label{fig:contour0}
\end{figure}

\section{Results: wind \& bubble morphology}
\label{sec:res-morph}
The result of an episodic explosive event at the centre of Milky Way would depend mainly on the rate of energy and mass input (and therefore on the SFR), the distribution of density through which the bubble ploughs its way (the disc and CGM gas density profile) and the epoch under consideration. 
 We fix these parameters based on the morphology of the resulting bubble, in light of the observed morphology of the FBs, and the emission properties. Therefore, we first discuss the morphology.

Figure \ref{fig:contour0} shows the colour-coded contours of density and temperature for our fiducial run, $\mathcal{L}= 5\times 10^{40}$ erg s$^{-1}$ at $t=27$ Myr (corresponding to $\approx 10^5$ supernovae over this time). 
 The snapshot clearly shows the structure of a standard  stellar wind scenario \citep{weaver1977}.  There is an outer shock (at a vertical distance of $\approx 10$ kpc), an enhancement of density in the shocked CGM/ISM and shocked wind region, near the contact discontinuity (at a vertical distance of $6\hbox{--}8$ kpc), as well as the inner free wind region (below a vertical distance of $\sim 6$ kpc).  The figure also shows a second reverse shock at height of $\sim 2$ kpc which arises because of the presence of two component density structure related to the CGM and the disc.

\begin{figure}
\includegraphics[trim=5.0cm 2.0cm 0.0cm 0.0cm, clip=true, width=6cm, height=10cm, angle=-90]{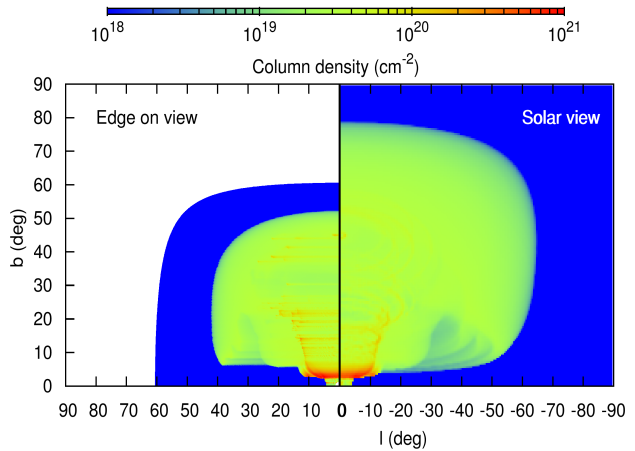}
 \caption {Snapshots of column density from edge-on position but without projection effects (left panel) and Solar vantage point with projection effects  (right panel),  for the same physical parameters as in Figure \ref{fig:contour0}. The boundary of our simulation box  (15 kpc) corresponds to  an angle $\sim 60^\circ$ from a distance of 8.5 kpc, and shows up in the left panel. }
 \label{fig:col_dens}
\end{figure}

Since we are at a distance of $8.5$ kpc from the centre of the Galaxy, and the wind-cone extends $\sim 4$ kpc at a height of $5-6$ kpc, much of the observed structure is influenced by geometrical projection effects.
Figure \ref{fig:col_dens} illustrates the idea by showing the map of column density as viewed from an edge-on vantage point from infinity, as well as its appearance from the point of view of the solar system. In order for the column density not to be dominated by the disc material, we have considered only the gas for which the total non-azimuthal speed $\left( \sqrt{v^2-v_\phi^2} \right)$ is larger than $20$ km s$^{-1}$. From the edge-on position, the Galactic coordinates are computed as $l=\tan ^{-1} (R/8.5 \, {\rm kpc}), \, b=\tan ^{-1} (z/ 8.5 \, {\rm kpc})$,\footnote{These formulae are valid only for $R,z \ll 8.5$ kpc , or equivalently $l,b \ll 45^\circ$.} whereas, from Solar view point (right panel), we have considered the projection effects accurately (see Appendix \ref{app:los} for details). In projection from the Solar system position, the bubble appears bigger in angular size. 
Note that we have used the axisymmetry property of our 2D simulations to get the projected maps presented in this paper.

 The difference between the left and right panels of Figure \ref{fig:col_dens} highlights the importance of taking projection effects into account when comparing the morphology of the simulated bubble with the observed FBs. With the projected column density map at hand, we can discuss the logic behind fixing the epoch of the phenomenon at $27$ Myr.

As explained below, the X-ray emission expected from the outer shock (shocked circumgalactic medium, CGM) is likely associated with the observed Loop-I feature in X-rays. This feature is also observed in soft $\gamma$-rays. The location of the outer shock depends strongly on the time elapsed, and helps us to fix the time at $27$ Myr. The radius of the outer shock in a constant luminosity-driven wind, according to \citet{weaver1977}, is given by ${\cal R} \approx ({\cal L}t^3/\rho)^{1/5}$ 
\be
\label{eq:ros}
\approx 10{\rm ~kpc} \left ( \frac{{\cal L}}{5\times 10^{40} {\rm erg~s}^{-1}}  \frac{0.001 m_p}{ \rho } \left [\frac{t}{27 {\rm Myr}} \right ]^3 \right )^{1/5},
\ee
matching the outer shock location in Figure \ref{fig:contour0}. Moreover, with this choice, we find that the location of the contact discontinuity matches the edge of the FBs. This indicates that the emission in different bands coming from the FBs is created within the the contact discontinuity. In addition, as shown below, the morphology of emission in different wavebands  remarkably matches the predictions based on this choice of time elapsed (namely, 27 Myr) and therefore, in turn, supports the idea that some part of the Loop-I feature is likely associated with the FBs. A point to note in Eq. \ref{eq:ros} is that the outer shock radius depends more sensitively on time rather than SFR or the CGM density. 

While Eq. \ref{eq:ros} is strictly valid only for an homogeneous and isotropic medium, and with isotropic energy injection, we expect it to be roughly valid, even with anisotropic AGN jets. Most AGN-based models consider a shorter age ($\sim 1$ Myr), which comes at the expense of a much larger mechanical power (up to  $10^{44}$ erg s$^{-1}$, \citealt{guo2012,yang12}). The velocity of the outer shock is given as ${\cal V} \approx 3 {\cal R}/5 t$
\be
\label{eq:vos}
\approx 200~{\rm km~s}^{-1} \left ( \frac{{\cal L}}{5\times 10^{40} {\rm erg~s}^{-1}} \frac{0.001 m_p}{ \rho } \left [ \frac{10 {\rm kpc}}{\cal R} \right ]^{2} \right)^{1/3},
\ee
comparable to the sound speed in the hot CGM ($\sim 180$ km s$^{-1}$), implying a weak shock in case of $\mathcal{L} = 5\times 10^{40}$ erg s$^{-1}$ as seen in Figure \ref{fig:contour0}. A more powerful AGN jet acting for 1 Myr with ${\cal L} \sim 10^{44}$ erg s$^{-1}$ will result in a very strong shock, ruled out by X-ray observations that show only a slight enhancement of temperature and density  across the FB edge as observed by \cite{kataoka2013}.

 Though we assume that the injection region is spherical symmetric, a departure from this assumption does not change the qualitative/quantitative picture much. The effect of different injection geometries has been discussed in section \ref{dis:geometry}.

\section{Results: emission in different wavebands}
\label{sec:res-emission}
We discuss the results of our calculation for the emission in different bands in this section, and compare with the observed features. Various emission mechanisms have been discussed in the literature for different bands -- gamma-rays, X-rays, microwave and radio, and most of the debate so far has centred around the $\gamma$-ray radiation mechanism (hadronic or leptonic), whether or not particles are being advected from the disc or accelerated {\it in situ}. However, among the emission in different bands, the X-ray emission from thermal gas suffers the least from any assumptions regarding accelerated particles and magnetic fields. We, therefore, discuss the X-ray emission first.

Since we have estimated the age of the Fermi bubbles to be 27 Myr (as discussed in the previous section), we perform detailed analysis for the fiducial run (S10; see Table \ref{table:runs}), at $t = 27$ Myr in this as well as in all the following sections.

\subsection{X-ray}
\label{subsec:xray}
Observationally, two limb-brightened X-ray arcs, called `northern arcs', are seen in the north-east quadrant adjacent to the FB. In the southern hemisphere, a `donut' feature is observed. Then there is the Loop-I feature extending up to $b \sim 80^\circ$ and from $50^\circ$ to $-70^\circ$ in longitude. The diffuse X-ray emission also shows a dip in intensity in the FB region  \citep{su2010}. Recently \cite{kataoka2013} have scanned the FB edge to look for differences in the X-ray brightness. They found that the temperature does not vary across the edge but there is a $50\%$ decrease of the emission measure (EM) when moving from outside to inside of the bubble. 

\begin{figure}
\centering
\includegraphics[trim=5.0cm 2.0cm 0.0cm 0.0cm, clip=true, width=6cm, height=10cm, angle=-90]{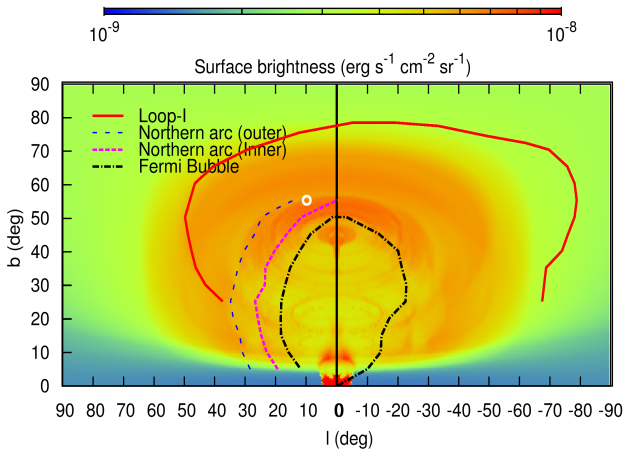}
\caption{Simulated X-ray emission map in $0.7\hbox{--}2.0$ keV band for the fiducial run (S10), over plotted with the observed edges of the Loop-I, northern arcs and the  northern FB. The white circle represents the region where we have compared the estimated emission measure with the observations mentioned in the text . } 
\label{fig:mekal}
\end{figure}

We show the surface brightness of X-ray emission from the simulated bubble in Figure \ref{fig:mekal}, in the $0.7\hbox{--}2.0$ keV band (ROSAT R6R7 band) considering the Mekal plasma model (from XPEC) for emission.  While calculating the X-ray surface brightness, we also consider the contribution from an extended CGM where the  hydrostatic state has been extrapolated to 100 kpc. \footnote{ For our isothermal CGM the density profile is centrally peaked so that surface brightness (SB) is dominated by the inner CGM, but halo gas extended out to 100 kpc makes a non-negligible contribution (50\%) compared to when it is confined to the 15 kpc box. The contribution of extended halo depends somewhat on the CGM density profile at large radii which is observationally unconstrained.}

 From Fig \ref{fig:mekal}, we find that (1) the diffuse emission has a dip at FB and extends to the Loop-I feature in the form of a parachute (this fixes the age of FBs in our model, as mentioned earlier),   and roughly delineates the feature leaving aside the slight asymmetry. (2) The location of the two arcs roughly matches the enhanced brightness (between $60^\circ < b < 50^\circ$, at $l\sim 0$). (3) The surface brightness is $ \approx 6 \times 10^{-9}$ erg s$^{-1}$ cm$^{-2}$ sr$^{-1}$, also consistent with observations \cite{su2010} who found  $\sim  2\times 10^{-8}$ erg s$^{-1}$ cm$^{-2}$ sr$^{-1}$. Although the contrast of the X-ray dip in the figure shown here is a bit less than observed,  we note that the final observed counts through an instrument will depend on the details of spectral modelling. It is clear from figure \ref{fig:contour0} that the gas inside the bubble has a temperature ($\sim 2\times 10^6$ K), lower than the shell temperature ($\sim 3.5 \times 10^6$). Therefore, the intensity when folded through an instrument to estimate counts will show a higher contrast. These simulated features reasonably match the observed structure in X-ray images.
\begin{figure}
\centering
\includegraphics[width=0.27\textheight, angle=-90]{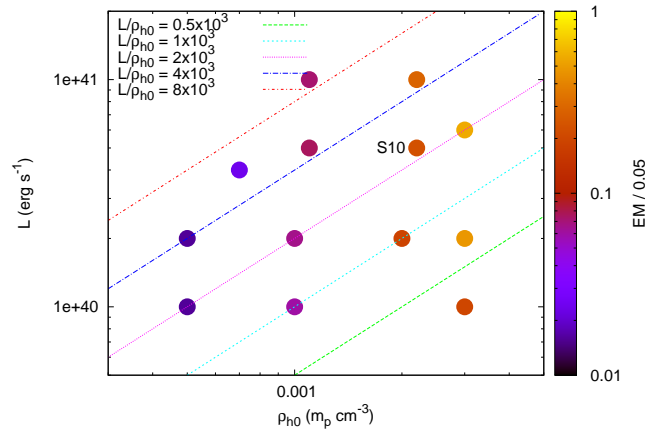}
\caption{Comparison of the estimated emission measure (EM) of the 'X-ray parachute' with the observed value $0.05$ cm$^{-6}$ pc. The filled circles represent the position of individual runs in parameter space (as mentioned in Table \ref{table:runs}) and the colour of each point represents the value of EM$/0.05$. Different values of $\mathcal{L}/\rho_{h0}$ in $10^{40}$ erg s$^{-1}$ m$_p^{-1}$ cm$^3$ have been shown by different straight lines. The fiducial run has been shown by 'S10' in this Figure.}
\label{fig:EM_comp}
\end{figure}
Since the intensity of the `X-ray parachute' mainly depends on the background CGM density, we use the emission measure (EM) of the parachute at $(l, b) \approx (10, 55)^\circ $ to match the EM of N1 point as observed by \cite{kataoka2013} (shown by the white circle in Figure \ref{fig:mekal}).  Figure \ref{fig:EM_comp} shows the estimated values of the emission measure (EM $\equiv \int n^2 dl$) for $0.24\hbox{--}0.38$ keV gas compared to the observed value of 0.05 cm$^{-6}$ pc for different runs (see Table \ref{table:runs}).  The figure shows that only for the central densities of $2\hbox{--}3.5 \times 10^{-3}$ m$_p$ cm$^{-3}$, the estimated EM is close to the observed value. This, along with the surface brightness of the `X-ray parachute' sets a constrain on the allowed background density of the CGM.

 The different straight lines  in figure \ref{fig:EM_comp} represent different values of $\mathcal{L}/\rho_{h0}$ (this ratio determines the radius of the outer shock for a given time; see Eq. \ref{eq:ros}). These lines of constant ${\cal L}/\rho_{h0}$ are also found to be crucial in determining the shape of the wind, and hence the projected shape within the contact discontinuity. For lower values of $\mathcal{L}/\rho_{h0}$,  the opening angle of the wind is much smaller than observed in FB (we assume that the gamma-rays of FBs come from the free and the shocked wind, as we discuss later). For larger values of $\mathcal{L}/\rho_{h0}$, though the opening angle matches with the base of the FB, the extent of the wind (in $l$) at high latitudes exceed the observed width of the bubble. However, the shapes arising from the runs lying on $\mathcal{L}/\rho_{h0} = 2\times 10^{3} \times 10^{40}$ erg s$^{-1}$ m$_p^{-1}$ cm$^3$ line have maximum similarity with the observed FB shape. 

Therefore, the constraint on $\rho_{h0} \left(\approx 2-3.5 \times 10^{-3}\right)$ from X-rays and the requirement for the FB shape leave us with a small parameter space in figure \ref{fig:EM_comp} which implies $\mathcal{L} \approx 5\hbox{--}7 \times 10^{40}$ erg s$^{-1}$. Since modelling of thermal X-rays is least uncertain as compared to the non-thermal radio and gamma-ray emission, we consider $\mathcal{L} = 5\times 10^{40}$ erg s$^{-1}$ and $\rho_{h0} = 2.2 \times 10^{-3}$ m$_p$ cm$^{-3}$ for our Galactic wind model parameters to calculate microwave and $\gamma$-ray emission.

\subsection{Microwave Haze}
\label{subsec:haze}
Microwave observations ($23$ GHz, with {\it WMAP and Planck}; \citealt{dobler08,planck2013}) show emission from $|b| \lesssim 35^{\circ}$  region on either side of the plane, termed the `microwave haze'. Diffuse radio emission is also seen in the 408 MHz map \citep{haslam1982} where the emission traces the Loop-I feature.  The 23-70 GHz emission spectrum shows a spectral index $\beta = 2.56$ (brightness temperature $T_b \propto \nu^{-\beta}$) which indicates the presence of an electron spectrum of spectral index $x = 2.2$ \citep{planck2013}. 
The $2.3$ GHz observation also reveals  polarised lobes and ridges in both hemispheres. The polarisation level in the ridges is $25\hbox{--}31\%$. The ridges in the north-east quadrant coincides with the FB edge and the x-ray shells, and it is found that the magnetic field is aligned with the ridges \citep{carretti2013}. The low-frequency
emission extends westward beyond the FBs in both hemispheres and the spectrum 2.3-23 GHz spectrum becomes softer as we go away from the Galactic center.

In order to estimate the emission from relativistic particles, we assume that the particle (either hadrons or leptons) energy density is a fraction of the {\it total} energy density of the gas (internal or thermal energy as well as the energy density due to fluid motion) . This is expected in the case of internal shocks and turbulence in the gas, and due to {\it in situ} acceleration of particles from these shocks. As Figure \ref{fig:contour0} shows, the  strong shocks that are likely to accelerate particles are traced by the shocked wind material which is the region inside the contact discontinuity (CD). Therefore, in order to trace the {\it freshly produced} accelerated particles, we use a tracer in the simulation that tracks the injected wind material from where most of the microwave/gamma-rays are emitted. Also, in order to avoid the disc material along the line of sight, we discard from our analysis the gas with a non-azimuthal velocity less than 20 km s$^{-1}$.

We can estimate the microwave emission in our model assuming synchrotron emission and that the cosmic ray (CR) energy density is given by $u_{\rm cr} = \epsilon_{\rm  cr} \,u_{\rm gas}$, where $u_{\rm gas} = u_{\rm th} + u_{\rm kin}$ is the total energy density of  the gas as discussed above. The CR electron energy density is assumed to be $u_{\rm cr,e} = 0.05\, u_{\rm cr}$ as expected from the ratio $\left( m_e/m_p \right)^{(3-x)/2}$ for $x = 2.2$ (see, e.g., \cite{persic2014}). This fixes the electron spectrum, $n(E) dE = \kappa E^{-x} dE$, where the normalisation constant $\kappa$  is given by $\kappa=u_{\rm cr,e} (x-2)/(m_e c^2)^{2-x}$ (assuming a lower cut-off of Lorentz factor $\sim 1$). The synchrotron emissivity, in the presence of a magnetic field $B$ in the optically-thin limit, is (Eq. 18.18 in \cite{longair}). 
\begin{eqnarray}
\frac{\varepsilon_\nu^{syn}}{\hbox{erg s}^{-1} \hbox{cm}^{-3}  \hbox{Hz}^{-1}} &=& 1.7\times 10^{-21} a(x)\kappa B^{\frac{x+1}{2}}\nonumber\\
&\times& \left( \frac{6.26\times 10^{18} \, {\rm Hz}}{\nu}\right)^\frac{x-1}{2} \,,
\end{eqnarray}
where, $a(2.2) \approx 0.1$.  
For the magnetic field, we assume that the magnetic energy is also a fraction of the thermal energy and is given as $u_{\rm B} = \epsilon_{B}\,u_{\rm gas}$.

We therefore have the volume emissivity per unit solid angle as
\begin{eqnarray}
\frac{J_\nu^{syn}}{\hbox{erg s}^{-1} \hbox{cm}^{-3}\hbox{Hz}^{-1} \hbox{sr}^{-1}} &=& 2.6 \times 10^{-20} \, \epsilon_{\rm cr} \epsilon_{B}^{0.8}\, p^{1.8} \nonumber \\ 
&\times & \left( \frac{23 \, \hbox{GHz}}{\nu} \right)^{0.6}\,.
\end{eqnarray}
where, we have taken $u_{\rm gas} = (3/2) \,p$, and $p = p_{\rm th} + 1/3\, \rho v^2$ is the total pressure (thermal plus kinetic).
\begin{figure}
\centering
\includegraphics[trim=4.5cm 2.0cm 0.0cm 1.0cm, clip=true, width=6cm, height=9.5cm, angle=-90]{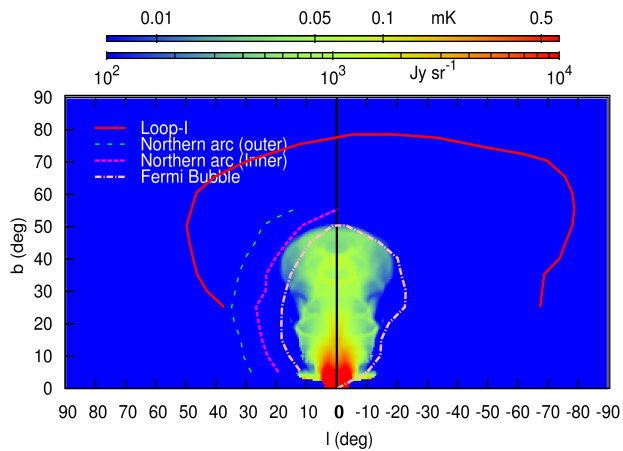}
\caption{23 GHz synchrotron emission (surface brightness) map for $\mathcal{L} = 5\times 10^{40}$ erg s$^{-1} $ and $\rho_{h0} = 2.2\times 10^{-3}$ m$_p$ cm$^{-3}$ with $\epsilon_{\rm cr} = 0.15$ and $\epsilon_{B} =  0.4$. The upper colourbar shows the brightness temperature in mK and the lower colourbar shows the brightness in units of Jy sr$^{-1}$.}
\label{fig:microwave}
\end{figure}

After calculating the surface brightness at 23 GHz from the FBs with $\epsilon_{\rm cr} = \epsilon_{B} = 1.0$, we found it to be approximately 15 times larger than the observed value of $800$ Jy sr$^{-1}$. This implies that 
\begin{equation}
\epsilon_{\rm cr}\,\epsilon_B^{0.8} \approx 1/15 \,.
\end{equation}
Note that we have an independent constrain on $\epsilon_{cr}$ because these same particles will also
emit $\gamma$-rays. Assuming $\epsilon_{\rm cr} = 0.15$, we get a constrain on the magnetic energy density that $\epsilon_B = 0.4$. This gives a magnetic field of strength $B = 3\hbox{--}5\,\mu$G considering $u_{\rm gas} \approx 0.7\hbox{--}3.0 \times 10^{-12}$ erg cm$^{-3}$ inside the bubble;  thus, 23 GHz emission
comes from electrons with $\gamma\approx 2\times 10^4$. Notice that our estimate of magnetic field is somewhat  lower than but consistent with other estimates \citep{su2010,  carretti2013, crocker2014b}.

The surface brightness of the 23 GHz emission is shown in Figure \ref{fig:microwave} and is consistent with observations. We also notice that the emission fills up the whole bubble volume which is consistent with recent observation. {\it Planck} has detected microwave emission from the whole FB region, although the intensity is rather small above $\sim b \ge 40^\circ$ ( see fig 9 of \cite{planck2013}), consistent with our results, given the uncertainties.

\begin{figure}
\centering
\includegraphics[trim=5.0cm 2.0cm 0.0cm 0.0cm, clip=true, width=6cm, height=9.5cm, angle=-90]{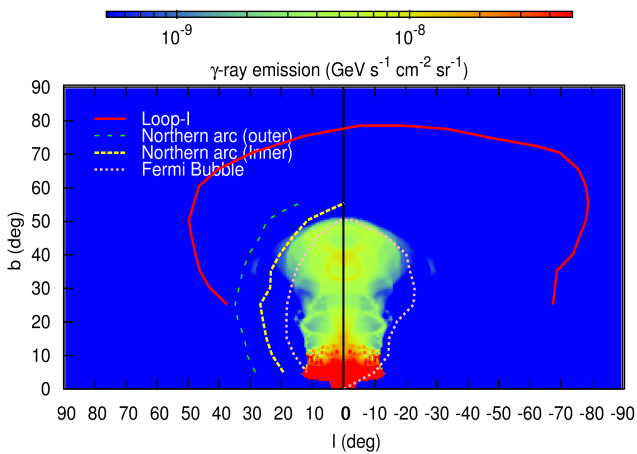}
\caption{Hadronic $\gamma$-ray emission map (surface brightness) as seen from the solar system location. Over plotted are the edges of the observed emission maps. }
\label{fig:gamma_hadronic}
\end{figure}
\subsection{$\gamma$-ray}
\label{subsec:gamma}
Observations show two $\gamma$-ray bubbles ($1.0\hbox{--}50.0$ GeV) on either side of the Galactic plane, being roughly symmetric about the plane. The northern bubble extends up to $b < 50^{\circ}$ and $|l| \lesssim 25^{\circ}$, which is almost same for the southern bubble. Another limb-brightened $\gamma$-ray feature extends up to  $80^{\circ}$ in $b$ and $\pm 70^{\circ}$ in $l$ in northern hemisphere and is known as Loop-I feature.
 The FB surface brightness is fairly uniform over the bubble and shows no limb brightening. The $\gamma$-ray spectrum of the FB is also flat ($dN/dE \sim E^{-2}$) and shows almost no softening with increasing height. The Loop-I feature, however, has a softer spectrum, $dN/dE \sim E^{-2.4}$, and this has led \cite{su2010} to conclude that Loop-I is a part of the disc and has no connection with  the bubble.

 In order to estimate the $\gamma$-ray emission from the simulated bubble, we consider two possible emission mechanisms, hadronic and leptonic. Below we discuss them in detail.
\subsubsection{Hadronic emission}
\label{subsubsec:hadronic}
In the hadronic model, cosmic ray (CR) protons 
undergo hadronic collisions with thermal gas protons and produce $\gamma$-ray via pion decay. The volume emissivity for this emission in the Fermi-LAT band (1GeV-100GeV) can be written as 
\begin{equation}
L^{pp} \simeq \frac{3}{2}\times \frac{1}{3}\times f_{\rm bol} \, u_{\rm cr,p} \, n_p \, \sigma_{pp}  \, \kappa_{pp} c\,,
\end{equation}
where, $n_p$ is the gas proton number density, $u_{\rm cr,p} \approx u_{\rm cr}$, is the CR proton energy density, $f_{\rm bol} \simeq 0.4$ is the fraction of the total luminosity that is emitted in the Fermi-LAT band. The factor $3/2$ corrects for the presence of heavy ions among the beam and target nuclei, 
$\sigma_{pp} = 4\times 10^{-26}$ cm$^{2}$ is the corresponding interaction cross-section and  $\kappa_{pp} = 0.5$ is the hadronic inelasticity \citep{crocker2014}. 

Figure \ref{fig:gamma_hadronic} shows the emission map in 1-100 GeV from hadronic collisions assuming $\epsilon_{\rm cr} = 0.15$ as discussed in the previous section. The surface brightness is fairly uniform and fills the observed region of FB. However, the average intensity is only $\lesssim 1\%$ of the observed value of $1.4\times 10^{-6}$ GeV s$^{-1}$ cm$^{-2}$ sr$^{-1}$ \citep{su2010} in this band.

\subsubsection{Leptonic}
\label{subsubsec:leptonic}
In the leptonic model, CMB photons are inverse Compton scattered by relativistic electrons and produce $\gamma$ rays.\footnote{We do not consider up-scattering of UV and IR photons because CMB dominates far away from the galactic disc. A proper modelling of the $\gamma$-ray emission due to interstellar radiation field (ISRF) is beyond the scope of this paper.}
The required CR electron energy is $1\hbox{--}100$ TeV to produce $\gamma$ rays of 1-100 GeV. Therefore, the corresponding Lorentz factors of the electrons range from $\gamma = 2\times 10^6$ to $2\times 10^7$. However, at such high $\gamma$ values the electron spectrum would likely suffer a cooling break  because of synchrotron and IC losses. Assuming a typical age of the electrons to be $t_{\rm age} = 1.5$ Myr, and considering the magnetic field to be $B = 4 \,\mu$G, as obtained in section \ref{subsec:haze},  we get  the Lorentz factor at the break to be $\gamma_b = 10^6$. Typically in a simple steady-state model for the evolution of the relativistic electron distribution function, the cooling break occurs at a $\gamma$ for which the cooling time equals the age. However, here we are considering time-dependent particle acceleration in turbulence and internal/termination shocks, and the effective age of electrons can be much shorter than the FB age.

\begin{figure}
\centering
\includegraphics[width=0.27\textheight, angle=-90]{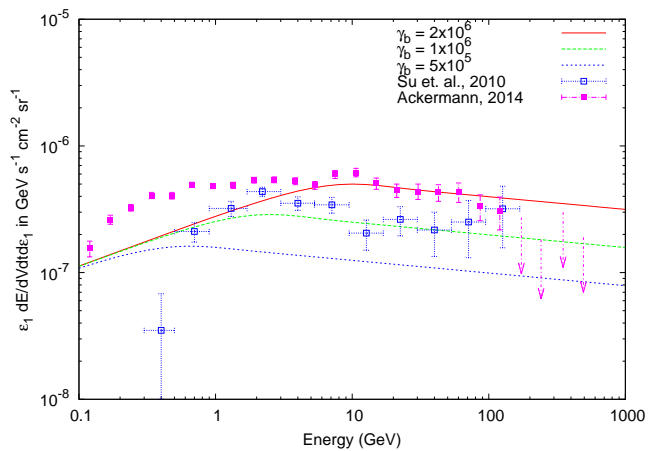}
\caption{Output spectra for leptonic $\gamma$-ray emission (green dashed line). The blue empty squares and the magenta filled squares show the observed data points \citep{su2010, ackermann2014},  and the green (dashed) line shows the spectra calculated by us for $\gamma_b = 10^6$. The plot also shows the spectrum for $\gamma_b = 5\times 10^5$ (blue dotted line) and $\gamma_b = 2\times 10^6$ (red solid line) for comparison. Notice that we do not consider a high energy cutoff for the electron distribution, which can account for the lack of gamma ray emission beyond few 100 GeV.}
\label{fig:lept-spect}
\end{figure}

Since $\gamma=10^6$ is close to the Lorentz factors needed for the leptonic emission to be in 1-100 GeV band, we consider a broken power law electron spectrum which has a spectral index $x_1 = 2.2$ (same as considered in synchrotron emission) below the break and the index drops by $\Delta x = 1$ after the break. The electron spectrum can be written as 
\begin{equation}
n(\gamma) = 
   \begin{cases}
         C\, \gamma^{-x_1} & \hbox{for}\,\, \gamma_l < \gamma \leq \gamma_b \\
         C\, \gamma_b^{x_2-x_1}\, \gamma^{-x_2}\,, & \hbox{for}\,\, \gamma_b < \gamma \leq \gamma_h 
   \end{cases}
\end{equation}
where, $x_2 = x_1 + 1$ is the spectral index after the break,  $\gamma_l$ and $\gamma_h $ are the lower and higher cut-off of the spectrum.  The normalisation factor $C$ can be written as 
\begin{eqnarray}
C &=& \frac{u_{\rm cr,e}}{m_e\,c^2}\,\left[ \frac{\gamma_l^{2-x_1}}{x_1-2} + \frac{\gamma_b^{2-x_2}}{x_2-2} \right]^{-1} \nonumber \\
&\approx & \frac{u_{\rm cr,e}}{m_e\,c^2} \frac{x_1-2}{\gamma_l^{2-x_1}}\,.
\end{eqnarray}
The output power per unit volume at an energy $\epsilon_1$ can be calculated as (Eq. 7.28a in \cite{rybicki})
\begin{eqnarray}
& &\epsilon_1 \frac{dE}{dVdtd\epsilon_1} = \frac{3}{4} c \sigma_T C \,\epsilon_1\,\int d\epsilon \left( \frac{\epsilon_1}{\epsilon}\right) v(\epsilon) \nonumber \\ 
&\times & \left[  \int_{\gamma_l}^{\gamma_b}\gamma^{-x_1-2} f\left(\frac{\epsilon_1}{4 \gamma^2 \epsilon}\right)d\gamma + \gamma_b  \int_{\gamma_b}^{\gamma_h}\gamma^{-x_2-2} f\left(\frac{\epsilon_1}{4 \gamma^2 \epsilon}\right)d\gamma  \right]\,\,, \nonumber \\
&&
 \label{eq:gamma-spec1}
\end{eqnarray}
where, 
\begin{equation}
f\left( x \right) = 2x\log(x)+x+1-2x^2 , \,\,\, \hbox{for}\,\, 0 < x < 1 
\end{equation}
and for seed photons with a blackbody spectrum at temperature $T_{\rm cmb}$,
\begin{equation}
v( \epsilon) = \frac{8\pi}{h^3 c^3} \frac{\epsilon^2}{\exp \left( \epsilon / k_BT_{\rm cmb} \right) - 1}\,\,.
\end{equation}
Here we use $u_{\rm cr,e} = 0.05\times u_{\rm cr} = 0.05 \epsilon_{cr} u_{gas}$ with $\epsilon_{cr} = 0.15$ (same as considered previously for synchrotron emission). The lower and the higher cut-off Lorentz factors are taken to be $\gamma_l = 1$ and $\gamma_h = \infty$. 
Eqn. \ref{eq:gamma-spec1} can be numerically integrated to give the resulting spectrum, which is shown in Figure \ref{fig:lept-spect}. The spectrum shows a reasonable match with the spectra as observed by \cite{su2010} and \cite{ackermann2014}. The figure also shows that the spectrum is consistent with the observations for $\gamma_b$ ranging from $5\times 10^5$ to $2\times 10^6$ and therefore it is robust under small uncertainties in the magnetic field or age of the electrons. While the flux in 1-100 GeV decreases for a smaller $\gamma_b$, it can be boosted by the additional IC up-scattering of the ambient starlight.
\begin{figure}
\centering
\includegraphics[trim=4.0cm 2.0cm 0.0cm 0.0cm, clip=true, width=6cm, height=10cm, angle=-90]{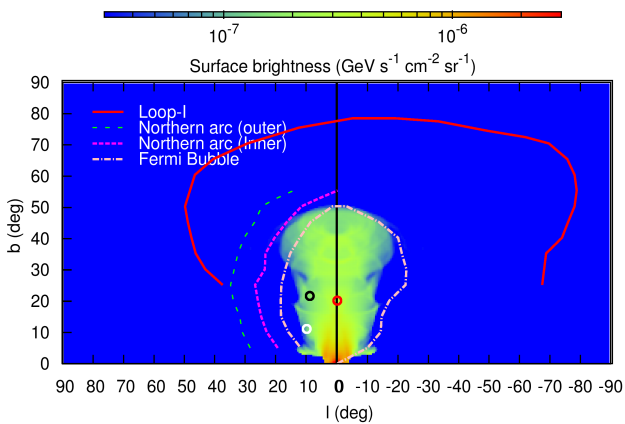}
\caption{Leptonic $\gamma$-ray emission map at 10 GeV as seen from the solar system location. The red and black circles represent the regions for which the velocity histograms have been been shown in Figure \ref{fig:vel-hist}. The white circle is the one where \citet{fox2014} have UV absorption data. }
\label{fig:gamma_leptonic}
\end{figure}

We also show the leptonic emission map at $10$ GeV in Figure \ref{fig:gamma_leptonic}. It shows a good match with the observed morphology of FBs. The surface brightness is also reasonably uniform over the region. Though the edge of the simulated bubble is not as smooth as observed, an introduction of magnetic field in the simulation can potentially make the bubble edge smoother.
\begin{figure}
\centering
\includegraphics[trim=2.0cm 2.0cm 0.0cm 0.0cm, clip=true, width=0.27\textheight, angle=-90]{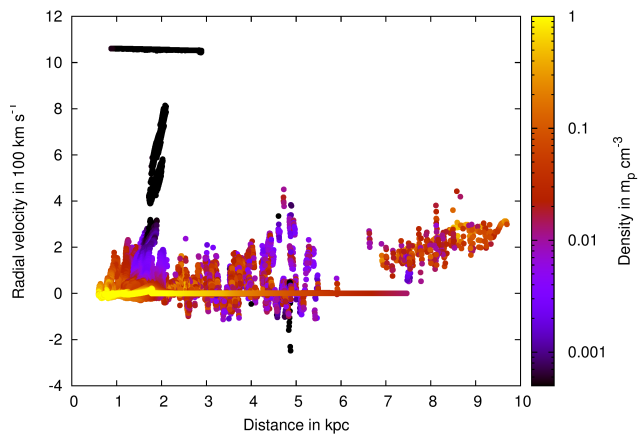}
\caption{Position-Velocity diagram of the gas parcels with $T < 2\times 10^5$ K. The colourbar represents the density of the gas parcels.}
\label{fig:RV}
\end{figure}
\section{Results: kinematics}
\label{sec:kinematics}
It is important to study the kinematics of FBs in order to infer  their origin, as it can give us crucial information about the speed of gas inside and around the bubbles.
 Recently, \cite{fox2014}  have detected ultraviolet absorption features in cold ($\sim 5\times 10^4$ K) and warm gas ($\sim 10^5$ K) phases at line of sight velocities of $-200$,  $+130$ and $+250$  km s$^{-1}$ towards quasar PDS 456 ($10.4^\circ, 11.2^\circ$). Using a simple model of biconical nuclear outflow, to obtain a line-of-sight velocity of $\sim -200$ km s$^{-1}$, they needed a cold/warm radial Galacto-centric outflow with  velocity ($v_{\rm gsr}$) $\gtrsim 900$ km s$^{-1}$. This is essentially because of the radial outflow assumption and the low inclination of the quasar sightline.

 The velocity structure in our simulated FBs has a more complicated structure than the simple models studied by \cite{fox2014}. In our simulation, the cold/warm clouds are formed by thermal and Rayleigh-Taylor instabilities at the 
interface of the injected gas with the CGM  (the contact discontinuity). The clouds formed at the conical surface of the contact discontinuity sometimes fall back due to the gravity (essentially a fountain flow; \citealt{shapiro1976}). However, the clouds at the top of the cone keep moving away from the centre because of the wind ram pressure. The low latitude cold/warm gas can have a wide angle and a large ($\sim 100$ km s$^{-1}$) line-of-sight velocity because the clouds are following non-radial trajectories (e.g., see the S2 sequence of clouds  in Fig. 12 of \citealt{sarkar15}).
\begin{figure*}
\centering
\includegraphics[trim=6.0cm 0.0cm 6.0cm 0.0cm, clip=true, width=0.4\textheight, height=0.3\textheight]{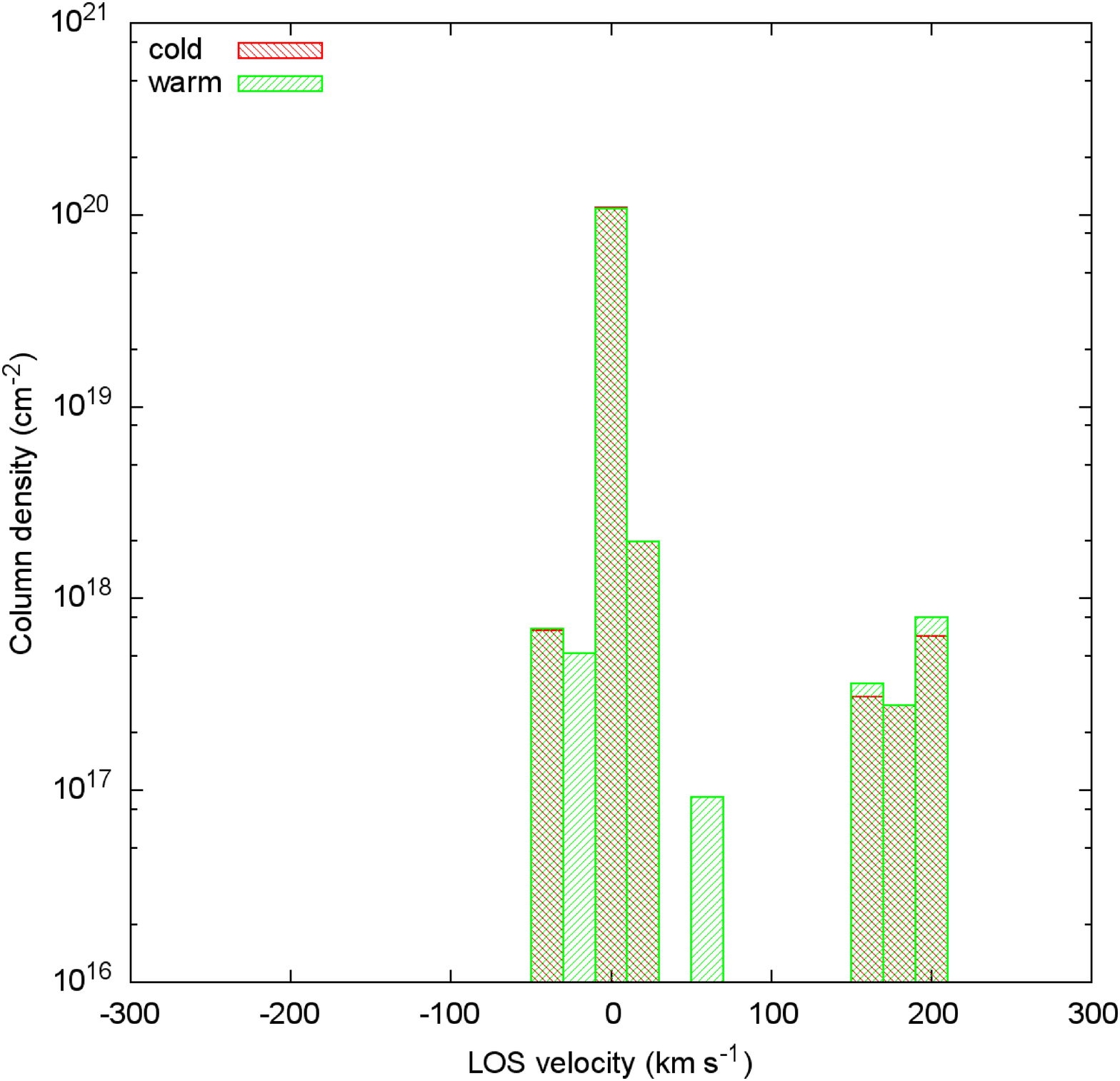}
\includegraphics[trim=6.0cm 0.0cm 6.0cm 0.0cm, clip=true, width=0.4\textheight, height=0.3\textheight]{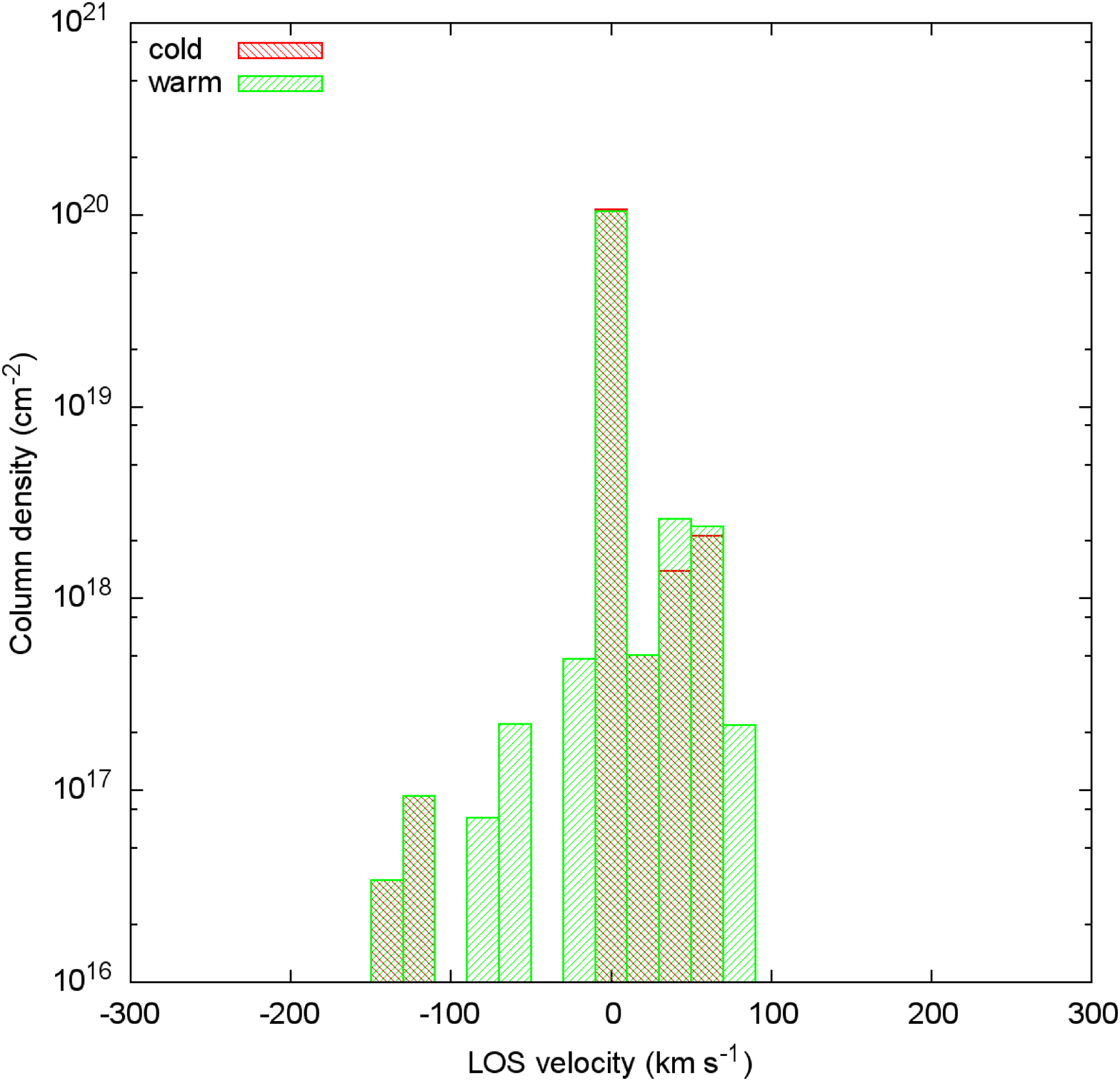}
\includegraphics[trim=6.0cm 0.0cm 6.0cm 0.0cm, clip=true, width=0.4\textheight, height=0.3\textheight]{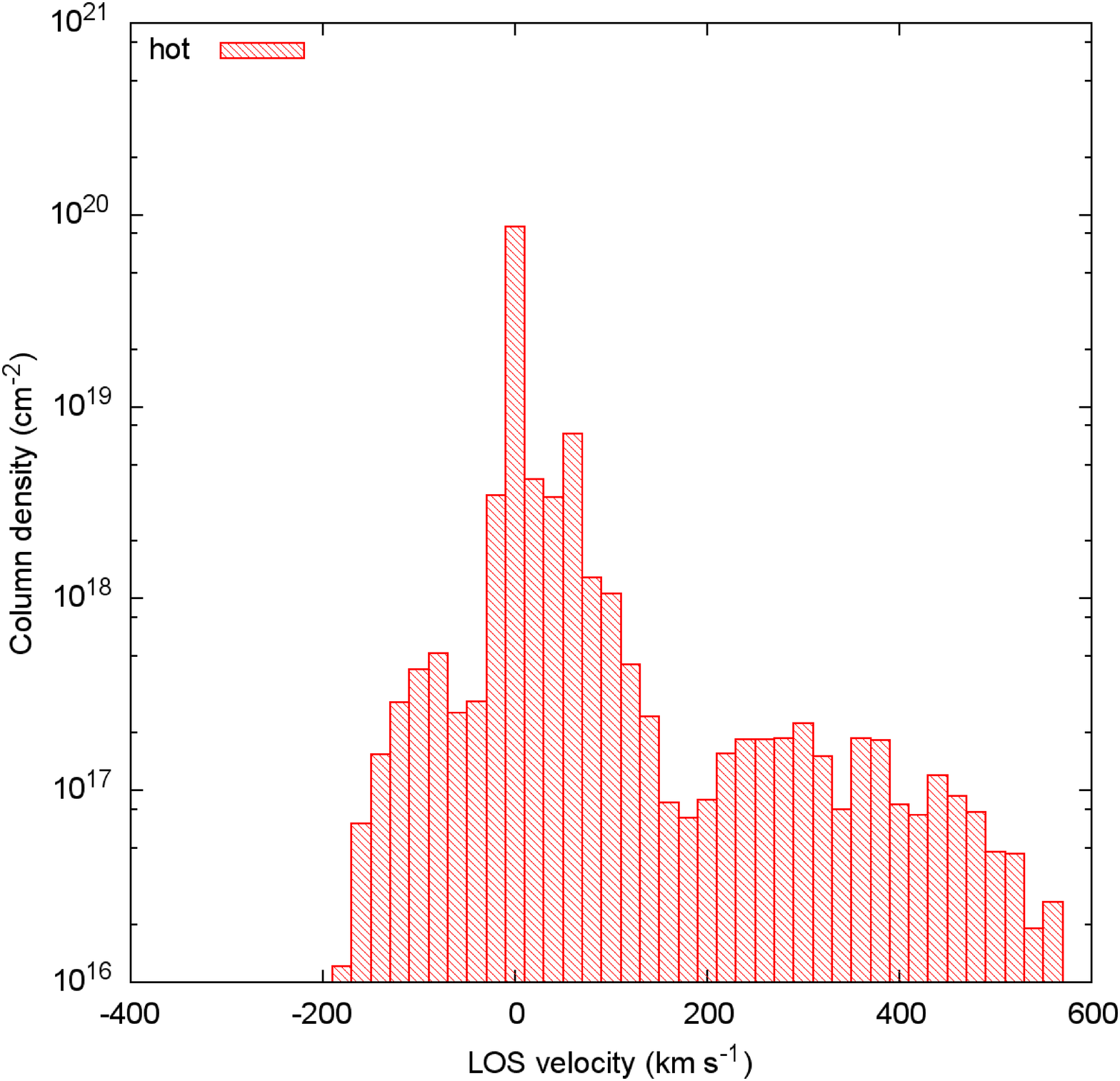}
\includegraphics[trim=6.0cm 0.0cm 6.0cm 0.0cm, clip=true, width=0.4\textheight, height=0.3\textheight]{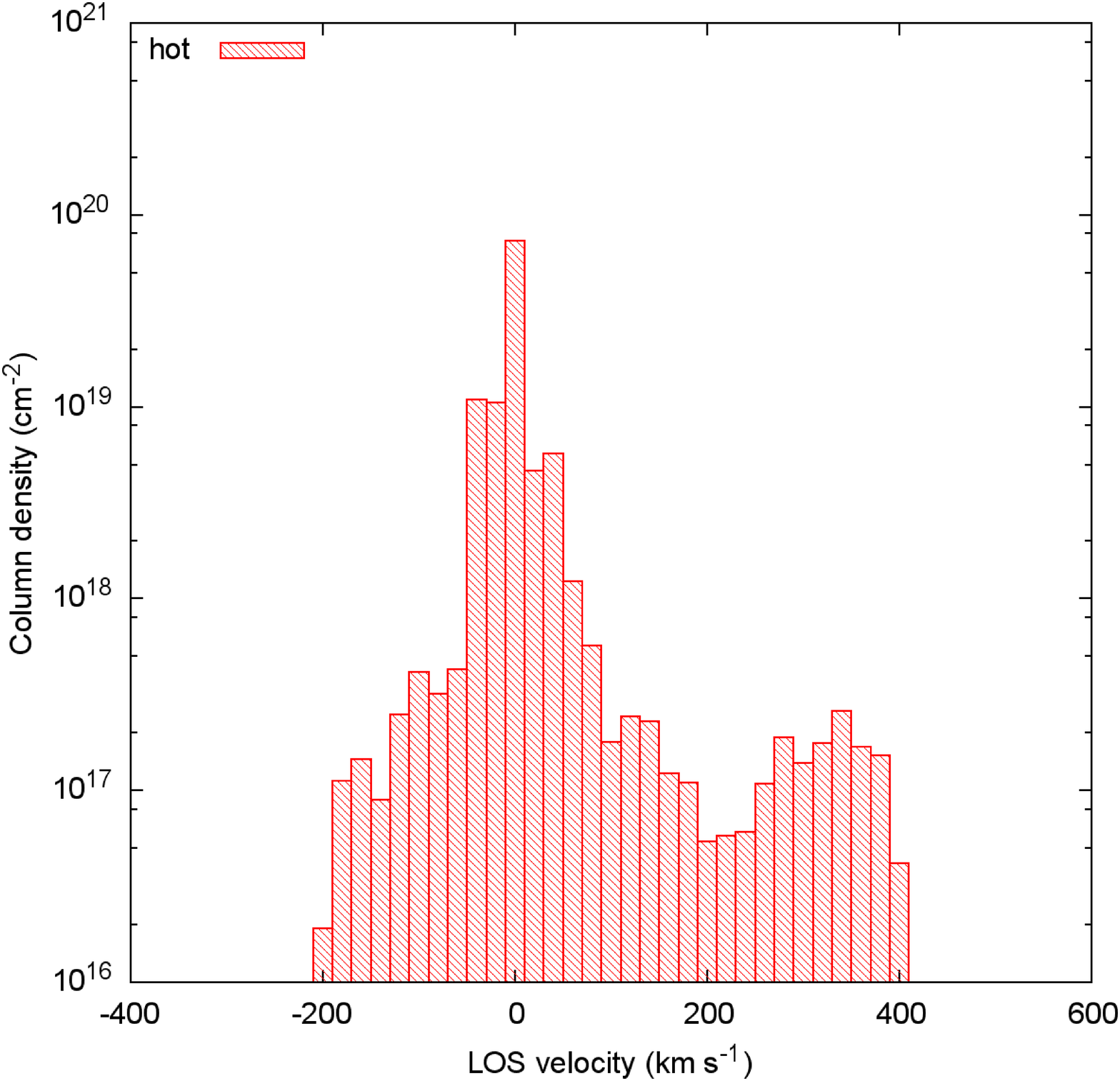}
\caption{Velocity histograms of the gas along ($0^\circ, 20^\circ$) in left panel and along ($9^\circ, 21.5^\circ$) in right panel. X-axis represents the line of sight velocity and the y axis represents the N(HII) corresponding to that velocity. The upper panel shows the velocity histogram for cold ($T< 4\times 10^4$ K) and for warm ($4\times 10^4 < T < 2\times 10^5$ K), whereas. the lower panel shows for the hot gas ($T > 10^6$ K). The LOSs have been pointed out in Figure \ref{fig:gamma_leptonic}.}
\label{fig:vel-hist}
\end{figure*}

 Figure \ref{fig:RV} shows the position-velocity diagram as seen from the Galactic centre. It shows that along with the positive velocity components of the warm gas, there are gas parcels which have negative velocities extending up to $-100$ km s$^{-1}$. This infalling gas can contribute to the negative velocities as observed by \cite{fox2014}.  

In figure \ref{fig:vel-hist}, we show the line-of-sight velocity ($v_{\rm los}$) histograms of the cold, warm and hot gas along two different lines-of-sights (shown by the black and red circles in Figure \ref{fig:gamma_leptonic}) that pass through the FB. We take into account the solar rotation velocity of $v_{\phi,\odot} = 220$ km s$^{-1}$ for this calculation (see Appendix \ref{app:los} for more details).  The central peak in all the histograms represent the stationary disc and halo gas. The upper panels clearly show that the $v_{\rm los}$ for the cold and warm gas can reach up to $-150$ km s$^{-1}$ and $+200$ km s$^{-1}$. 
We also show the velocity-histograms of the hot ($T>10^6$ K) gas in the lower panel of Figure \ref{fig:vel-hist}. Though the hot gas velocities extend all the way from $-200$ to $+600$ km s$^{1}$ for these two line of sights (LOS), the other LOSs show hot gas velocities extending from -$500$ to $700 $ km s$^{-1}$. Notice that though the hot gas in our simulation has high velocity ($\sim 1000$ km s$^{-1}$) within the free wind region,  the shape of the histograms differ from that of the high velocity gas considered by \cite{fox2014}. This is essentially because of the non-radial flow of hot gas induced by Kelvin-Helmholtz instability.

The above-mentioned results show that the kinematic signatures of our model  are consistent with observations and one does not necessarily need cold/warm gas with velocities $\gtrsim 900$ km s$^{-1}$.

\section{Discussion}
\label{sec:discussion}
\subsection{Diffusion of CRs}
In this paper, we have so far tracked CRs by using a tracer which confine them within the contact discontinuity (CD). In reality, CRs diffuse beyond the CD and produce extended $\gamma$-ray and synchrotron emissions. Since turbulence is not expected to be significant outside the CD, the {\it in-situ} acceleration of CRs becomes ineffective and the emission  contains a signature of ageing in its spectrum. We can estimate this length scale over which the relativistic electrons   diffuse before they lose their energy, by considering the diffusion coefficient to be \citep{gabici07}
\begin{equation}
D(E,B) = 10^{28}\,\left( \frac{E}{10 GeV} \right)^{1/2}\,\left( \frac{B}{3 \mu G} \right)^{-1/2}\,.
\end{equation}
For electrons of energy $E \sim 1$ TeV moving in a magnetic field of  $B \sim 10 \,\mu G$ (estimated from fluid compression at the forward shock), the length-scale of diffusion in $t_{\rm age} = 1.5$ Myr (see \S 4.3.2) is $\sigma = \sqrt{6 D t_{\rm age}} \approx 1.3$ kpc. 
This implies that CR would diffuse up to $\sim 10^\circ$  beyond the CD at a height of $8$ kpc. This extended emission can appear as 2.3 GHz radio lobe as observed by \cite{carretti2013}.
\subsection{Note on Loop-I}
\label{subsec:loopI}
The accelerated relativistic particles from the outer shock may also produce gamma-ray emission at the shock position. However, because of the absence of further acceleration mechanisms (\textit{viz.} turbulence) behind the shock, the particles will lose their energy and may have a spectrum that is different from that of FB. In case of leptonic emission at the outer shock, the electrons lose their energy after $t_{\rm age} = 1.5$ Myr giving rise to a faint gamma ray shell which can appear as a diffuse emission when viewed from the Solar vantage point. This may partially explain the observed  emission from Loop-I feature.

Incidentally, we note that recent observations by \cite{ackermann2014} (Fig. 13) have revealed a Southern counterpart of the Loop-I feature. This lends an additional support for the connection between the Loop-I and the FBs.
\subsection{Kinematics of the cold/warm gas}
The speed of the cold/warm gas is complex in nature as shown in Figure \ref{fig:RV}, ranging from $-150$ to $+1000$ km s$^{-1}$. The density of the clouds, however, decreases with increasing velocity making them hard to detect. Moreover, very high velocity clouds may appear to be moving with low LOS velocity because of our vantage point. Therefore,  even if star formation at the Galactic centre produces  very high velocity clouds (VHVC),  several factors  can  make them undetectable as VHVCs from the Solar vantage point. The $+1000$ km s$^{-1}$ streak in the figure represents adiabatically expanding free wind. Because of its cone-like geometry, the velocity dispersion along a line-of-sight can become large enough for a given column density  that any absorption feature corresponding to the free wind may not be visible.  
\subsection{Effects of the  injection geometry and CGM rotation}
\label{dis:geometry}
 Although  we have considered the injection region to be spherically symmetric with radius $r_{\rm inj} = 60$ pc, star formation can occur in a region of complicated geometry as observed by \cite{molinari12}. To understand the effect of different injection geometries, we have carried out the following set of runs for our fiducial value:  
\begin{itemize}
\item spherically symmetric, with $r_{\rm inj} = 40$ pc.
\item spherically symmetric, with $r_{\rm inj} = 100$ pc.
\item axisymmetric (about $R = 0$ axis) disc-like injection region, with a radius $R = 110$ pc and midplane to edge height $h = 42$ pc \citep{lacki14}.
\item axisymmetric ring-like injection injection region, with inner radius $R_{\rm in} = 70$ pc, outer radius $R_{\rm out} = 100$ pc and midplane to edge height $h = 50$ pc.
\item spherically symmetric, with $r_{\rm inj} = 60$ pc, same as the fiducial run but this time we have considered that the halo gas is also rotating with a speed equal to 10\% of the stellar rotation speed at $z = 0$ at that $R$.
\end{itemize}

 The other parameters, namely $\epsilon_{\rm cr}, \epsilon_{B}, \gamma _b (=10^6)$, have been kept fixed to those values mentioned in the text for the fiducial run. The projected leptonic $\gamma$-ray emission maps at 10 GeV for all the cases (at $t = 27$ Myr) have been shown in figure \ref{fig:inj_geo}. The figures show that apart from  a slight corrugation of the FB edge, the morphologies and intensities match quite well with each other. Therefore, 
the conclusions in this paper are not affected significantly by our assumptions regarding the geometry of the star formation region and halo rotation. 
\subsection{Effects of CR and magnetic pressure}
 In our model, the magnetic and the CR energy densities are, respectively, $0.4$ and $0.15$ times the gas energy density.  The addition of these forms of energy would surely increase the total energy content and hence the pressure inside the bubble. More so, because CRs and magnetic fields do not suffer significant radiation loses. Therefore, for the same bubble energetics  the required SNe energy injection, and consequently, the SFR will decrease approx. by a factor of 1.5 if we consider a strong coupling of these nonthermal pressures with the gas. This can bring down the required SFR to $0.3-0.4$ M$_{\odot}$ yr$^{-1}$. 
\section{Summary}
\label{summary}
In this paper, we have presented the results of  numerical simulations of SNe driven outflows in our Galaxy,  taking into account a gaseous disc ($T \simeq 10^4$ K) and halo gas ($T = 2.5 \times 10^6$ K), in order to explain the origin of the Fermi Bubbles and multi-wavelength features related to it.  We injected continuous SNe energy at the Galactic centre  for $50$ Myr. We assumed {\it in situ} acceleration of   relativistic particles  inside the contact discontinuity.  Our model can explain the gamma-ray emission and microwave haze as coming from the interior of the contact discontinuity via leptonic  and synchrotron emission respectively. In addition,  X-ray is  emitted  mostly by the shocked CGM. Given our vantage point at the Solar position, we considered the projection effects properly to calculate the morphology of the bubble. In order to understand the dynamics of the bubble, we have also studied the cold, warm and hot gas kinematics.

In our analysis,  we have used the observed surface brightness in different bands in order to constrain  the  background CGM density, the star formation rate at the Galactic centre and the magnetic field inside the bubble, self consistently.  We summarise the main results of our paper below. 
\begin{itemize}
 \item  The X-ray emission appears to have a parachute-like structure with a dip in intensity within the boundaries of  the Fermi Bubbles. The surface brightness of the parachute is comparable to the observed value only if the central CGM gas density is taken to be  $ 2\hbox{--}3.5 \times 10^{-3} \,\mbox{m}_p \,\mbox{cm}^{-3}$ and an extended CGM up to $100$ kpc. Considering the morphological aspects of the Fermi Bubbles along with the CGM, the injected mechanical luminosity is found to be $ 5\hbox{--}7 \times 10^{40}$ erg s$^{-1}$ which corresponds a SFR $\approx 0.5\hbox{--}0.7 $ M$_{\odot}$ yr$^{_1}$.  
 \item Assuming that a relativistic electron population of spectral index $x = 2.2$  gives rise to the microwave haze via synchrotron emission, we  estimated the magnetic field inside the bubble to be $3\hbox{--}5\,\mu$G which is little lower but consistent with other estimations. This electron population  diffuses out from the contact discontinuity  and produces polarised radio emission as observed. 
\item Considering the above constrained CGM, SFR and the magnetic field, the $\gamma$-ray emission from the region inside contact discontinuity appears to have the shape and brightness comparable to observations. 
\item The speed of the cold ($T< 4 \times 10^4$ K) and warm ($4\times 10^4 < T < 2\times 10^5$ K) clumps can vary from $- 150$ km s$^{-1}$ to $+200$ km s$^{-1}$ (warm), whereas, the hot ($T > 10^6$ K) gas have a higher dispersion in their velocities which range from $\sim - 500$ to $+700$ km s$^{-1}$.  The kinematics of the cold/warm clumps  appear to have the characteristics that are similar to  recent observations. While \citet{fox2014} argue for a large radial velocity outflow (e.g., associated with AGN-driven outflows) because of a small angle between the radial direction and the line of sights through low latitudes, Figure \ref{fig:vel-hist} shows that we can obtain line-of-sight velocities consistent with observations because of non-radial trajectories and the negative radial velocities of entrained cold clumps without requiring large radial velocity of cold/warm gas.
\end{itemize}

To conclude, our work shows that star formation  at the Galactic centre  can give rise to the observed Fermi Bubbles and the multi-band structures related to it.  Furthermore,  modelling  these structures  can yield the basic Galactic parameters  such as the hot CGM density and magnetic field and opens up a new window to study  high energy interactions in the Milky Way.

\section*{Acknowledgements}
We would like to thank Dipankar Bhattacharya, Peter Biermann, Roland Crocker, Nazma Islam, Jun Kataoka, Uri Keshet, Preeti Kharb, Guo-bin Mou and Biswajit Paul for useful discussions. We also thank anonymous referee for  constructive comments. This work is partly supported by the DST-India grant no. Sr/S2/HEP-048/2012 and an India-Israel joint research grant (6-10/2014[IC]). KCS is partly supported by the CSIR-India (grant no. 09/1079(0002)/2012-EMR-I).

\footnotesize{


\appendix
\section{Projection effects}
\label{app:los}
In our analysis, we consider the Cartesian coordinates whose origin is at the GC and the variables are calculated from the local standard of rest (LSR) and in terms of galactic coordinates ($l\,,b$).  

Figure \ref{fig:vlos_dia} explains the working geometry.  For a given ($l\,,b$), the unit vector along the line of sight can be given as
\begin{equation}
\hat{\mathbf{L}} =  -\hat{\mathbf{i}}\cos(b)\cos(l) - \hat{\mathbf{j}} \cos(b)\sin(l) + \hat{\mathbf{k}}\sin(b)\,.
\end{equation}
\begin{figure}
\begin{center}
\includegraphics[trim = 0.0cm 25.0cm 0.0cm 0.0cm, clip=true,  width=8.0cm, height=7.0cm]{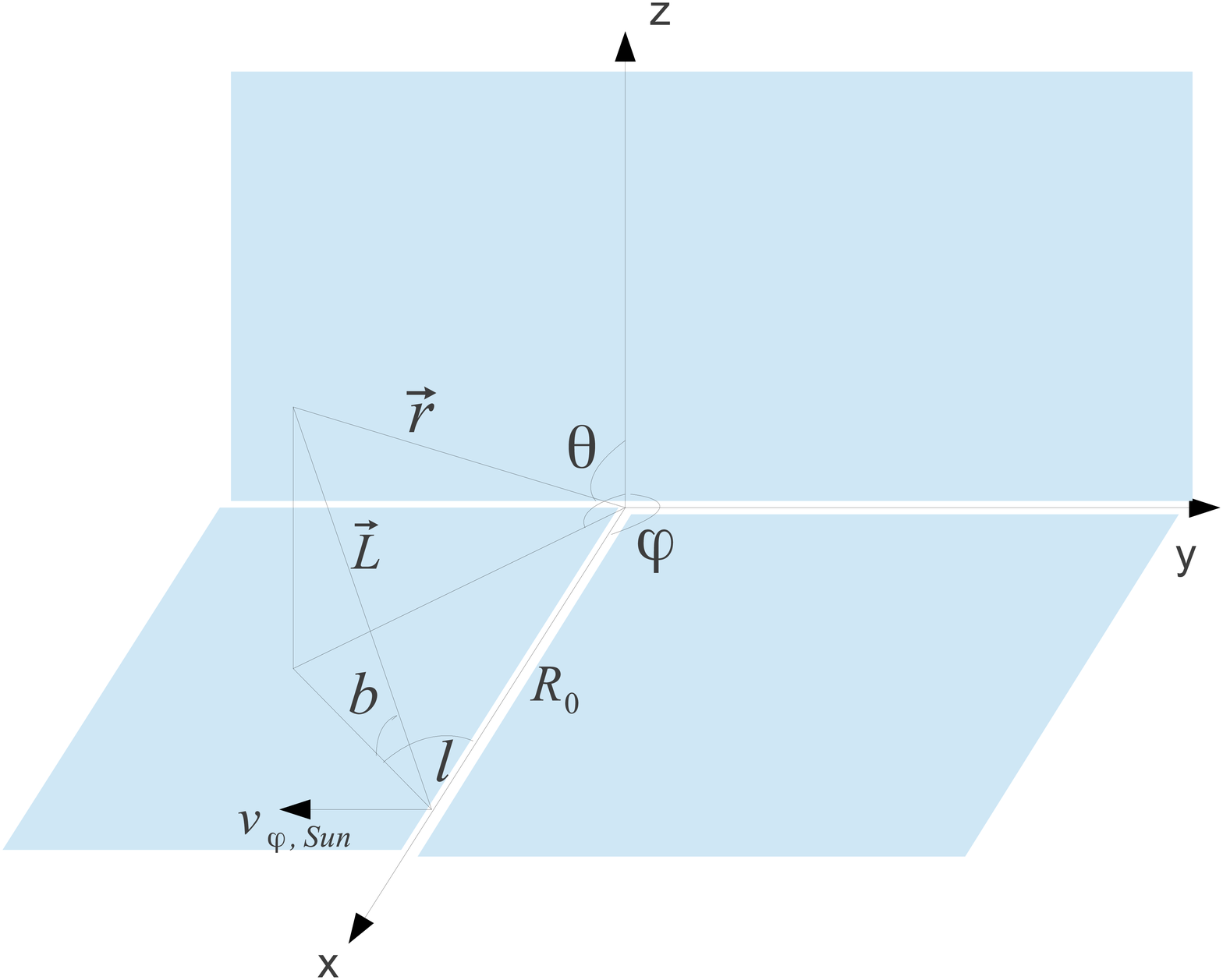}
\end{center}
\caption{The geometry for calculating the projected maps and LOS velocity from the Solar vantage point. The coordinate system has its origin at the Galactic center (GC), whereas the Sun is situated at a distance of $R_0 = 8.5$ kpc from the GC along the X-axis. The Y-Z plane represents the plane perpendicular to the Galactic disc. The Galactic coordinates are marked as $l$ and $b$, the simulation coordinates are shown as $r$ and $\theta$ in the diagram and $L$ is the distance of the point under consideration from the Sun.}
\label{fig:vlos_dia}
\end{figure}
\begin{figure*}
\centering
\includegraphics[trim=4.5cm 2.0cm 0.0cm 3.0cm, clip=true, width=0.3\textheight, height=0.4\textheight, angle=-90]{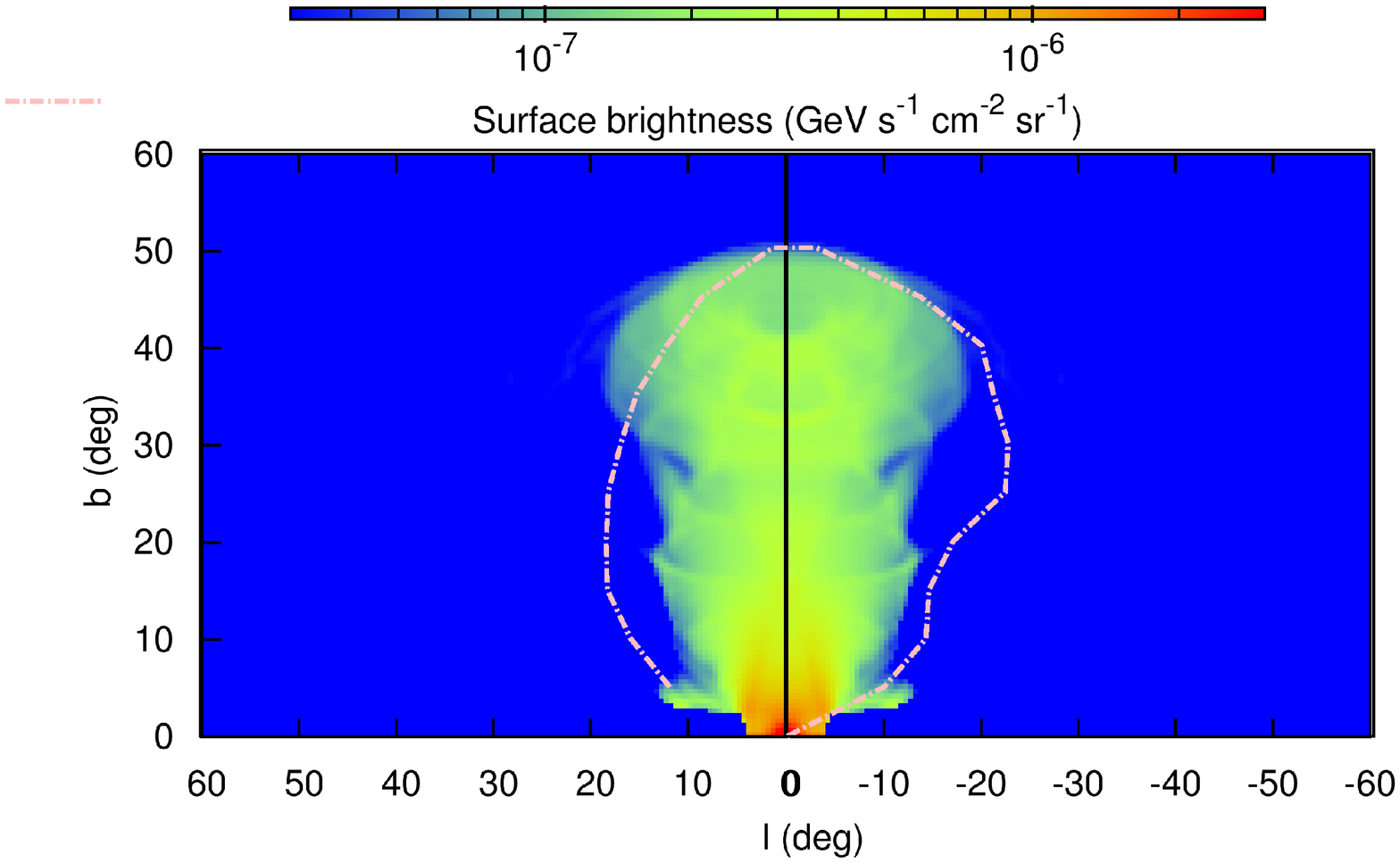}
\includegraphics[trim=4.5cm 2.0cm 0.0cm 3.0cm, clip=true, width=0.3\textheight, height=0.4\textheight, angle=-90]{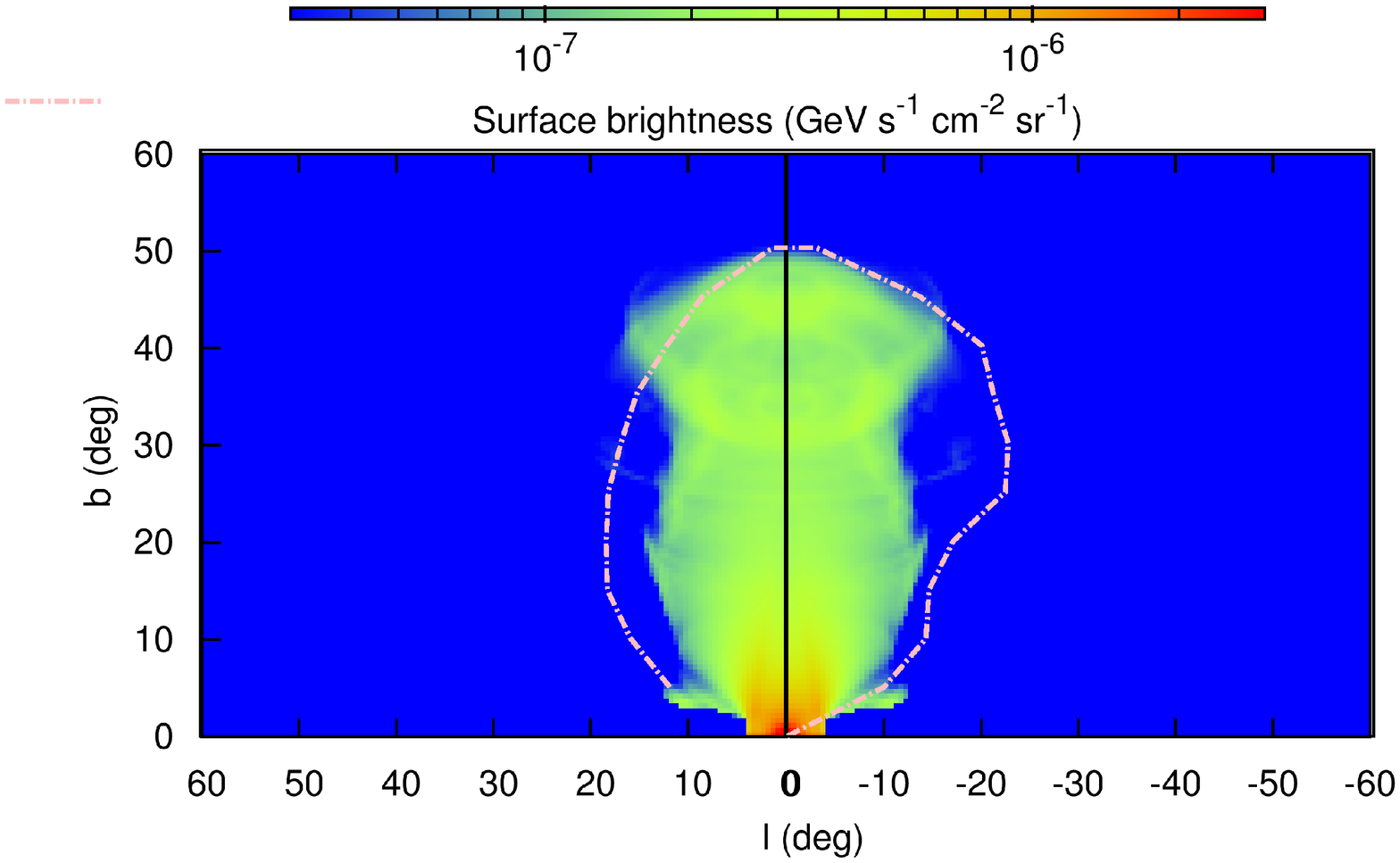}
\includegraphics[trim=7.2cm 2.0cm 0.0cm 3.0cm, clip=true, width=0.25\textheight, height=0.4\textheight, angle=-90]{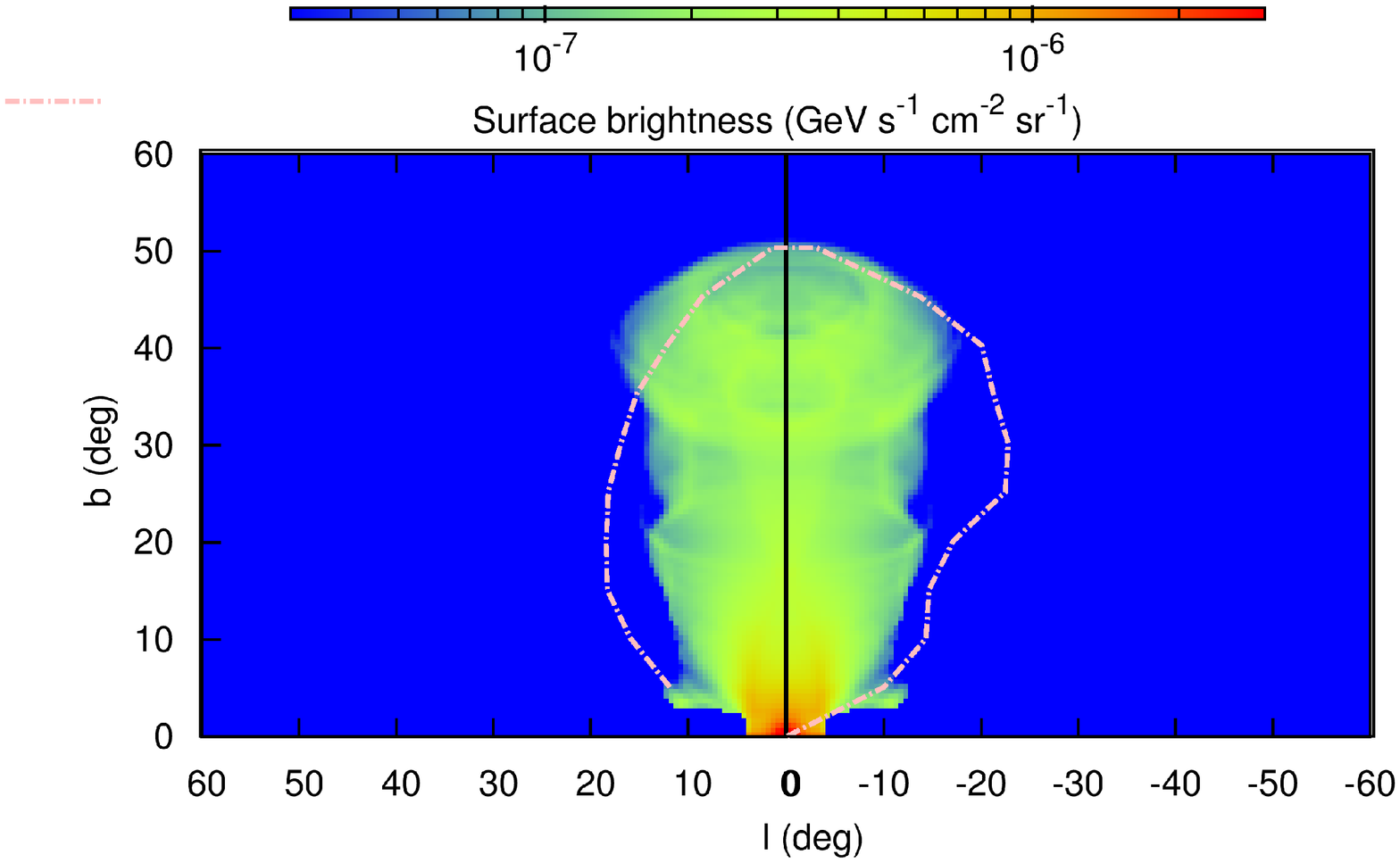}
\includegraphics[trim=7.2cm 2.0cm 0.0cm 3.0cm, clip=true, width=0.25\textheight, height=0.4\textheight, angle=-90]{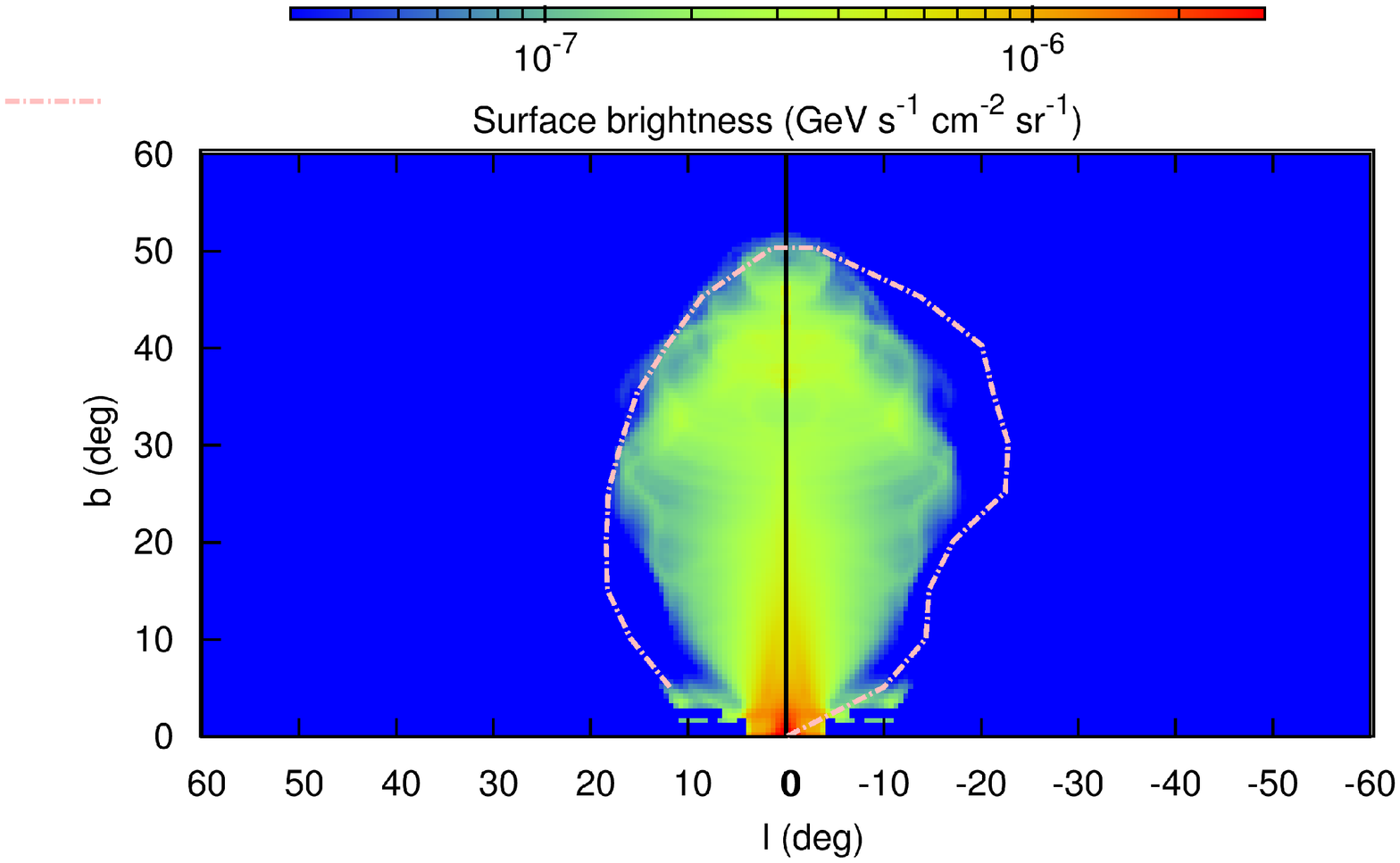}
\includegraphics[trim=7.2cm 2.0cm 0.0cm 3.0cm, clip=true, width=0.25\textheight, height=0.4\textheight, angle=-90]{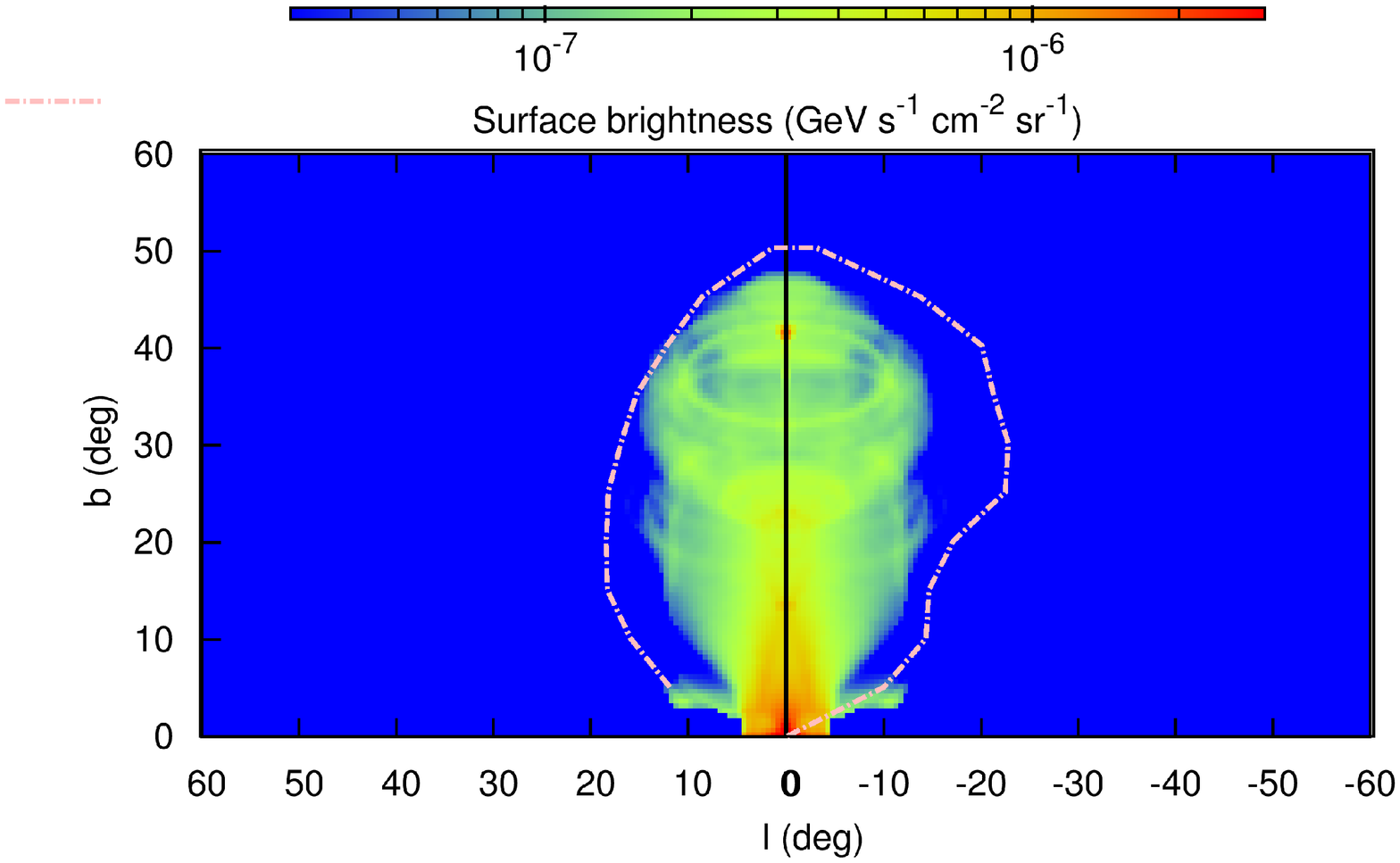}
\includegraphics[trim=7.2cm 2.0cm 0.0cm 3.0cm, clip=true, width=0.25\textheight, height=0.4\textheight, angle=-90]{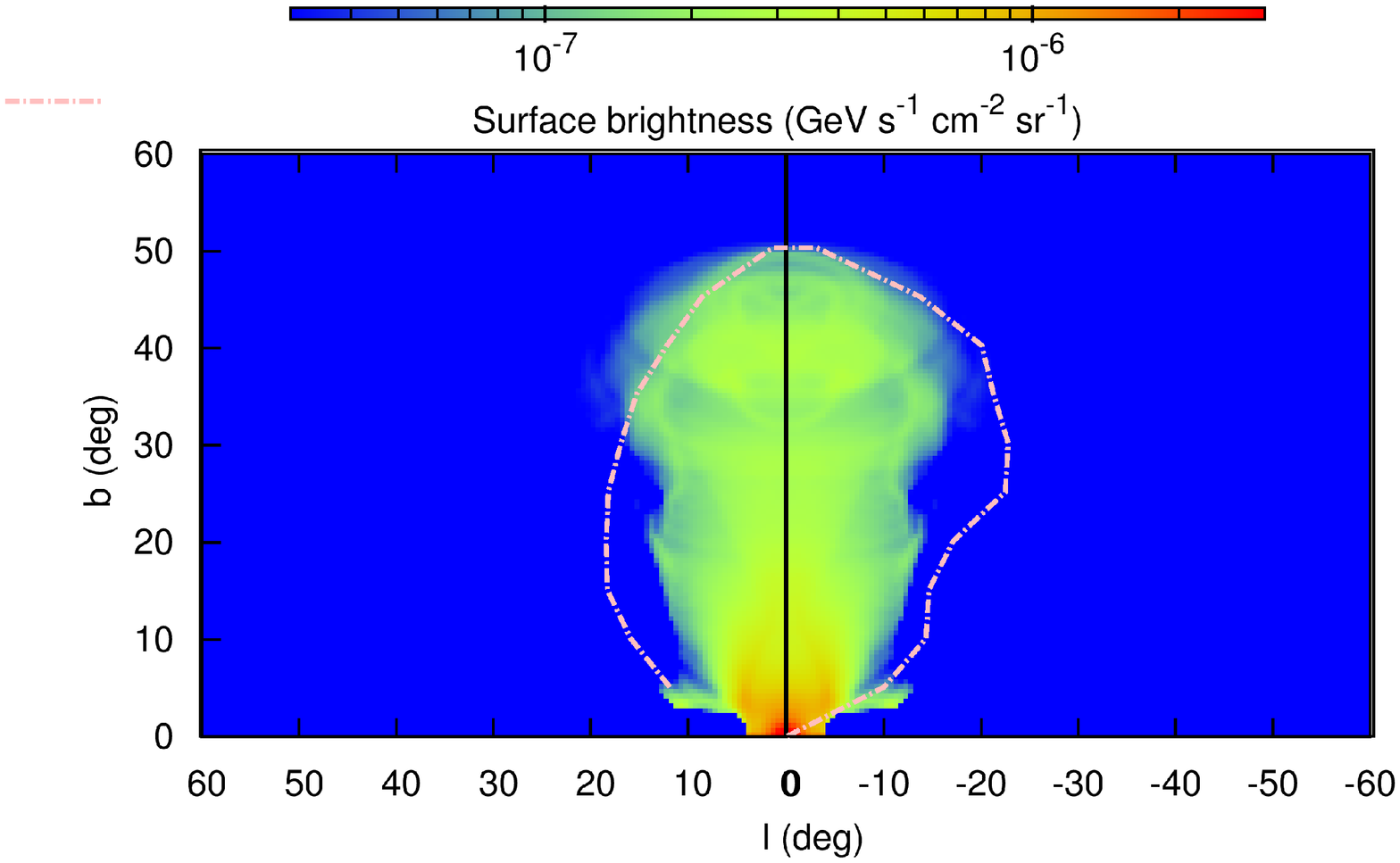}
\caption{ Effect of different injection geometry and CGM rotation for our fiducial run S10 (see Table \ref{table:runs}) on the $\gamma$-ray emission at 10 GeV. The geometries of the injection regions, from top-left to bottom-right, are i) spherically symmetric, with injection radius $r_{\rm  inj} = 60$ pc (the fiducial value), ii) spherically symmetric, with $r_{\rm  inj} = 40$ pc, iii)  spherically symmetric, with $r_{\rm  inj} = 100$ pc,  iv) axisymmetric, with disc-like injection zone, v) axisymmetric with a ring-like injection region and vi) spherically symmetric, with $r_{\rm inj} = 60$ pc and a rotating halo. The dot-dashed line shows the observed edge of the FB.}
\label{fig:inj_geo}
\end{figure*}

The unit vectors in spherical coordinates, used in the simulation, can be written in terms of the Cartesian unit vectors as 
\begin{eqnarray}
\hat{\mathbf{r}} &=& \hat{\mathbf{i}}\sin(\theta)\cos(\phi) + \hat{\mathbf{j}}\sin(\theta)\sin(\phi)      
                                   + \hat{\mathbf{k}}\cos(\theta) \nonumber \\
\hat{\mathbf{\theta}} &=&  \hat{\mathbf{i}} \cos(\theta)\cos(\phi) + \hat{\mathbf{j}}\cos(\theta)\sin(\phi) 
                                           -  \hat{\mathbf{k}}\sin(\theta) \nonumber\\
 \hat{\mathbf{\phi}} &=& - \hat{\mathbf{i}}\sin(\phi) + \hat{\mathbf{j}}\cos(\phi) \,.
\end{eqnarray}
Therefore, the component of the actual velocity, $\overrightarrow{v} = \hat{\mathbf{r}}v_r + \hat{\mathbf{\theta}}v_{\theta} + \hat{\mathbf{\phi}} v_{\phi}$, along the LOS is 
\begin{equation}
v_{\rm los} = \overrightarrow{v}.\hat{\mathbf{L}}\,.
\end{equation}
However, because of the solar rotation of $v_{\phi, \odot} $ on the plane, the actual LOS velocity along some ($l, b$) is 
\begin{eqnarray}
v_{\rm los} &=& \overrightarrow{v}.\hat{\mathbf{L}} - v_{\phi,\odot}\,\sin(l)\,\cos(b) \nonumber \\
      &=&- \cos(b)\cos(l) \Bigl [ v_r \sin(\theta)\cos(\phi) + v_{\theta}\cos(\theta)\cos(\phi)  - v_{\phi}\sin(\phi) \Bigr ]  \nonumber \\
      &-& \cos(b)\sin(l) \Bigl [ v_r \sin(\theta)\sin(\phi) + v_{\theta}\cos(\theta)\sin(\phi)  + v_{\phi}\cos(\phi) \Bigr ] \nonumber \\
      &+& \sin(b) \Bigl[ v_r \cos(\theta) - v_{\theta}\sin(\theta) \Bigl] - v_{\phi,\odot}\,\sin(l)\,\cos(b)\,.
\end{eqnarray} 

\end{document}